\documentclass[graybox, secnum]{svmult}

\usepackage{amsmath,amssymb}
\usepackage{mathptmx}       % selects Times Roman as basic font
\usepackage{helvet}         % selects Helvetica as sans-serif font
\usepackage{courier}        % selects Courier as typewriter font
\usepackage{type1cm}        % activate if the above 3 fonts are
\usepackage{makeidx}         % allows index generation
\usepackage{graphicx}        % standard LaTeX graphics tool
\usepackage{multicol}        % used for the two-column index
\usepackage[bottom]{footmisc}% places footnotes at page bottom
\usepackage{hyperref}        %for hyperlinks
\usepackage{soul}            % for high-lighting of text
\hypersetup{colorlinks=true,urlcolor=blue}
\usepackage{amsfonts}
\usepackage[square,numbers]{natbib}
\makeindex             % used for the subject index
\begin{document}
\newcommand{\ds}{\mathrm{d}}
 \newcommand{\bq}{\begin{equation}}
 \newcommand{\eq}{\end{equation}}
 \newcommand{\bqn}{\begin{eqnarray}}
 \newcommand{\eqn}{\end{eqnarray}}
 \newcommand{\nb}{\nonumber}
 \newcommand{\lb}{\label}
%\tableofcontents{}
\title*{Loop Quantum Cosmology: Physics of Singularity Resolution and its Implications}
\author{Bao-Fei Li$^{1,2}$  and Parampreet Singh$^2$\thanks{corresponding author}}
\institute{Institute for Theoretical Physics $\&$ Cosmology, Zhejiang University of Technology, Hangzhou, 310032, China.
%\email{libaofei@zjut.edu.cn}
\and Department of Physics and Astronomy, Louisiana State University, Baton Rouge, Louisiana 70803, USA %\email{psingh@lsu.edu}}
}

\maketitle

\abstract{The occurrence of singularities where spacetime curvature  becomes infinite and geodesic evolution breaks down are inevitable events in classical general relativity (GR) unless one chooses an exotic matter violating weak energy condition. These singularities show up in various physical processes, such as the gravitational collapse, the birth of the universe in the standard cosmology as well as the classical solutions of the black hole spacetimes.  In the last two decades, a rigorous understanding of the dynamics of quantum spacetime and the way it resolves singularities has been achieved in loop quantum cosmology (LQC) which applies the concepts and techniques of loop quantum gravity to the symmetry reduced cosmological spacetimes. Due to the fundamental discreteness of quantum geometry derived from the quantum theory, the big bang singularity has been robustly shown to be replaced by a big bounce. Strong curvature singularities intrinsic in the classical cosmology are generically resolved  for a variety of cosmological spacetimes including anisotropic models and polarized Gowdy models. Using effective spacetime description the LQC universe also provides an ultra-violet complete description of the classical inflationary scenario as well as its alternatives such as the ekpyrotic and matter bounce scenarios. In this chapter\footnote{Invited chapter to appear in the "Handbook of Quantum Gravity", edited by Cosimo Bambi, Leonardo Modesto and Ilya Shapiro, Springer (2023).}, we provide a summary of singularity resolution and its physical implications  for various  isotropic and anisotropic cosmological spacetimes in LQC and analyze robustness of results through variant models originating from different quantization prescriptions.  }

\section*{Keywords}

Loop quantum cosmology; Singularity resolution; Effective dynamics; Anisotropic models; Inflationary paradigm; Ekpyrotic scenario; Matter-bounce scenario

\section{Introduction}
\renewcommand{\theequation}{1.\arabic{equation}}\setcounter{equation}{0}\textbf{}

Theorems of Penrose, Hawking and Geroch show that singularities, such as the big bang, are the generic features of spacetimes in general relativity (GR)  \cite{Geroch:1968ut,Hawking:1970zqf,Hawking-Ellis}. These are the boundaries of spacetime at which the classical evolution stops and the known laws of physics reach the limit of their validity. A new theory is needed to go beyond these final boundaries of classical spacetime to make definitive predictions about the initial state of our universe, to understand the emergence of the classical space and time,
and to explore the new physics of the very early universe, and, deep inside the black hole interiors.
It has been long believed that an understanding of quantum gravitational effects would provide
important insights on the resolution of singularities. One such avenue, where quantum gravity
effects have been used to understand the problem of singularities is loop quantum cosmology (LQC)  \cite{Ashtekar:2011ni}.
 It is a non-perturbative canonical quantization of homogeneous spacetimes using the techniques
of loop quantum gravity (LQG) \cite{rovelli-book,Thiemann:2007pyv,pullin-book}, a candidate theory of quantum gravity based on Ashtekar-Barbero variables.
Though LQG is not yet a complete theory, in the last three decades  sufficient mathematical control has
been achieved to use it to quantize spacetimes with reduced degrees of freedom, such as cosmological
and black hole spacetimes.
A key prediction of LQG is that geometrical operators have discrete eigenvalues, at least at
the kinematical level. However, due to its complexity,  it becomes much easier and more straightforward  to test the technical and conceptual issues of LQG in the context of symmetry reduced spacetimes where great simplifications resulting from symmetry reduction lead to a manageable quantum theory. Taking into account the influx of the observational data of the early universe with increasing precision in the recent years, cosmology provides an ideal platform where predictions from the quantum theory are expected to be tested by direct observational signals. In this setting, LQC provides a novel for quantizing cosmological spacetimes using the infrastructure of LQG.

LQC inherits discrete quantum geometry from LQG. This results in significant qualitative differences between LQC and earlier attempts to quantum cosmology, such as the Wheeler-DeWitt theory. These qualitative
differences are most prominent only when spacetime curvature is large and comparable to the Planck
scale. At small spacetime curvatures, classical differential geometry turns out to be an excellent
approximation of the discrete quantum geometry of LQC, and GR is recovered. LQC thus passes an
important test required for any theory of quantum cosmology/gravity – it has the correct low energy
limit. Since LQC is a canonical quantization, dynamics is governed by a Hamiltonian constraint,
which due to the underlying quantum geometry turns out to be quantum difference equation. The
structure of the Hamiltonian constraint in LQC is in a striking contrast to the one in the Wheeler-DeWitt theory where dynamics is governed by a differential Wheeler-DeWitt equation. Unlike the
Wheeler-DeWitt theory where the problem of singularities persists even after quantization, in LQC
discrete quantum geometry results in a quantum bounce when spacetime curvature reaches Planck
scale \cite{Ashtekar:2006rx,Ashtekar:2006uz,Ashtekar:2006wn}.

The first rigorous model of LQC with a physical Hilbert space was constructed  for a spatially-flat homogeneous and isotropic FLRW universe sourced with a massless scalar field  \cite{Ashtekar:2006rx,Ashtekar:2006uz,Ashtekar:2006wn}.  Its quantum theory was formulated on the lines of Dirac quantization approach for the constrained systems.  In comparison to the older quantum cosmology proposed initially by DeWitt \cite{DeWitt:1967yk}, LQC has the following two notable properties. First, it utilizes a quantum representation based on the holonomies of  the Ashtekar-Barbero connection and gauge-fixed  triads  which is unitarily inequivalent to the Schr\"odinger representation used in the Wheeler-DeWitt quantum cosmology. The second main difference arises from the underlying quantum geometry. In the geometric representation, the differential Wheeler-DeWitt equation is replaced by a quantum difference equation where the difference in geometric sector is determined by the quantum geometry.
The Hamiltonian constraint yields a Schr\"odinger-like equation, providing a dynamical evolution of physical states,  with temporal role played by the massless scalar field. While the latter is not discretized, the evolution generator captures the underlying quantum geometry and is a discrete operator.
In particular, the minimal non-zero eigenvalue of the area operator in LQG is incorporated in defining a non-local curvature operator \cite{Ashtekar:2003hd}, leading to the well-known $\bar \mu$ scheme in LQC \cite{Ashtekar:2006wn}. Using  sophisticated numerical simulations of the quantum difference equation \cite{Ashtekar:2006wn,Diener:2014mia,Singh:2018rwa}, it has been rigorously shown that irrespective of the initial conditions, the big bang singularity is resolved and replaced by a quantum bounce. The bounce takes place at a fixed maximum energy density in the Planck regime as is predicted by an exactly solvable model in LQC \cite{Ashtekar:2007em}. If one starts evolution of the universe when it is macroscopic and peaked on a classical GR trajectory at late times, it turns out that states remain sharply peaked across the bounce \cite{Corichi:2007am, Kaminski:2010yz}. On the other side of the bounce, there exists a contracting branch asymptoting to the same classical universe as in the expanding branch. The occurrence of the quantum bounce is a direct result of the minimal area gap of the loop over which holonomies are constructed. Using consistent histories formalism, one can show that the probability of bounce turns out to be unity \cite{Craig:2013mga}. In contrast, in the Wheeler-DeWitt theory, the expectation value of the volume operator follows the classical trajectories which inevitably end up with a vanishing volume signifying the encounter with the big bang singularity \cite{Ashtekar:2006wn}. Even if one considers an arbitrary superposition of expanding and contracting branches in the Wheeler-DeWitt theory the consistent quantum probability for bounce is zero and for big bang is unity \cite{Craig:2010ai,Craig:2010vf}. Therefore, two quantum theories, namely LQC and Wheeler-DeWitt theory,  only converge in the classical limit. In the Planck regime, they lead to distinct physical predictions.

Interestingly, with the help of a complete set of the  Dirac observables and the  well-defined semi-classical physical states, the quantum evolution of the universe in LQC is shown to be endowed with a continuum spacetime description which is governed by an effective Hamiltonian constraint. This effective dynamics captures the leading-order corrections of the quantum theory and yields modified Friedmann and Raychaudhuri equations in LQC \cite{Willis_thesis,Taveras:2008ke}. The effective dynamics has been extensively used in the literature and it plays an important role in demonstrating the robustness of the singularity resolution \cite{Singh:2009mz,Singh:2011gp,Singh:2014fsy,Saini:2017ipg,Saini:2017ggt,Saini:2016vgo} and understanding the Planck-scale physics of the LQC universe extended from some well-known scenarios developed in the classical cosmology, such as the inflationary models \cite{Singh:2006im,Xiong:2007ak,Ashtekar:2009mm,Corichi:2010zp,Mielczarek:2010bh,Ashtekar:2011rm,Ranken:2012hp,Artymowski:2012is,Gupt:2013swa,Corichi:2013kua,Linsefors:2013cd,Bonga:2015kaa,Ashtekar:2015dja,Ashtekar:2016wpi,Zhu:2017jew,Shahalam:2018rby,Sharma:2018vnv,Gordon:2020gel,Motaharfar:2021gwi}, ekpyrotic and the matter bounce scenarios \cite{,Bojowald:2004kt,Cailleteau:2009fv,Wilson-Ewing:2012lmx,Cai:2014zga,Haro:2017mir,Li:2020pww}. Extensive work has been performed in the last decade to understand quantum geometry effects on cosmological perturbations (see \cite{Agullo:2016tjh} for a review, and next chapter in this volume \cite{pert-chapter}).

With the success achieved by LQC in yielding a physically viable theory of quantum cosmology  in the simplest settings, the same techniques have been extended to loop quantize other cosmological spacetimes with more complicated structures, such as FLRW universe with spatial curvature \cite{Ashtekar:2006es,Szulc:2006ep,Corichi:2011pg}, Bianchi universe with anisotropies and spatial curvature \cite{Chiou:2007sp,Ashtekar:2009vc,Ashtekar:2009um,Wilson-Ewing:2010lkm}, Gowdy models with continuous degrees of freedom \cite{Martin-Benito:2008eza,MenaMarugan:2009dp,Brizuela:2009nk,Garay:2010sk,Martin-Benito:2010vep,Martin-Benito:2010dge}. The rigorous construction of the mathematical structure of the loop quantized cosmological spacetimes, including the kinematical Hilbert space, the Hamiltonian constraint operator, the Dirac observables as well as the physical inner product,  leads to a consistent picture of singularity resolution and the Planck-scale physics in each model. Various unique properties of the particular quantum spacetimes have been studied in detail by using the effective dynamics, resulting in a deeper understanding of how the quantum geometry effects can change the classical description of the cosmological spacetimes. For example, from the modified Friedmann and Raychaudhuri equations quantum geometry effects can be interpreted as making gravity repulsive in the Planck regime and resulting in a singularity resolution without any violation of any energy conditions.

Furthermore, attempts has  been made in the direction of incorporating more features from LQG into the LQC universe. Since the loop quantization in LQC is implemented in the symmetry reduced mini-superspace, its relationship with the cosmological sector of the full theory is not yet established \cite{Brunnemann:2007du,Engle:2007qh}  and ongoing research  is being conducted in order to reach a better understanding of the essential features of the LQG cosmology.   Recent progresses have been reported in the quantum-reduced loop quantum gravity  \cite{Alesci:2013xd}, the group theory cosmology \cite{Gielen:2013kla,Gielen:2013naa}, the path integral approach \cite{Han:2019feb}
and modified LQC models using the coherent states \cite{Yang:2009fp,Dapor:2017rwv}. %and employing the gauge covariant fluxes \cite{Liegener:2019ymd,Liegener:2019dzj}
In particular,  strong singularities are proved to be resolved and replaced by a quantum bounce as well in the modified LQC models \cite{Saini:2018tto} and moreover richer structures beyond standard LQC, such as an asymmetric bounce and a pure quantum state in the contracting phase and etc \cite{Li:2018opr}, have been observed which imply that the cosmological sector from LQG can be much more complicated   than expected. These new features of the modified LQC models have so far been  studied only in the simplest spatially-flat FLRW universe and their impacts on  other cosmological spacetimes with spatial curvature and anisotropies are still to be investigated.

In this chapter, we summarize some of the main results  of LQC on the singularity resolution for various types of cosmological spacetimes and discuss  resulting physical implications of quantum gravitational effects on the non-singular evolution of the LQC universe. The goal of this chapter is to not provide an exhaustive review of various techniques one may use and various questions one may pose to understand the nature of singularity resolution\footnotetext[1]{One example is to pose the question with respect to unitary evolution and loss of determinism. See for eg. \cite{Pawlowski:2014fba}.}, but only to give a flavor of the physical implications resulting from underlying quantum geometry assuming the validity of effective spacetime description in LQC. For completeness and to make the material accessible to a broader audience we also discuss some aspects of classical Hamiltonian cosmology in metric variables and the nature of  singularities.  The topics in each section are organized as follows. In Sec. \ref{Hamiltonian_cosmology}, the Hamiltonian formulation of GR is reviewed in the metric as well as  the Ashtekar-Barbero variables. The classical Hamiltonian constraints and the corresponding dynamical equations for the spatially-flat FLRW universe and the Bianchi-I universe are also discussed briefly.  In Sec. \ref{Sec:nature_of_singularities}, the types and the strength of the classical singularities that can be encountered in the cosmological spacetimes are addressed. For the spacetimes with anisotropies, we also review different shapes of singularities that arise from the anisotropic behavior of the directional scale factors when a singularity is approached. In Sec. \ref{sec:isotropic_model}, we go through in some detail the construction of the kinematic Hilbert space and the quantum Hamiltonian constraint operator in the spatially-flat, homogeneous and isotropic LQC universe. The properties of the resulting quantum difference equation are discussed along with the viability and the uniqueness of the improved dynamics (the $\bar \mu$ scheme). Sec. \ref{sec:effective_dynamics} is devoted to a summary of the effective dynamics for the $k=0$ homogeneous and isotropic LQC universe and its phenomenological applications. The effective dynamics, whose validity is assumed in all regimes, is extensively used in showing the generic resolution of the strong singularities and investigating the extensions of the inflationary, ekpyrotic and matter bounce scenarios in the LQC universe.  In Sec. \ref{quantization-anisotropies_inhomogenities},  we move onto the loop quantization of spacetimes with anisotropies and inhomogeneities.  In particular, we summarize the novel features of the loop quantization of Bianchi-I and Gowdy models and some of their physical implications on the inflationary paradigm.  In Sec. \ref{sec:alternative_cosmological_model}, we discuss different variants of LQC, such as the  modified LQC models. Here emphasis is on their distinct physical predictions as compared with those from standard LQC.   Finally, Sec. \ref{sec:conclusions} is attributed to a summary of the chapter.

\section{Hamiltonian cosmology}
\lb{Hamiltonian_cosmology}
\renewcommand{\theequation}{2.\arabic{equation}}\setcounter{equation}{0}

In this section we provide an overview of the Hamiltonian formulation of general relativity (GR) which serves as the first step to understand the main procedures of loop quantization of cosmological spacetimes in loop quantum cosmology. We start with the Arnowitt-Deser-Misner (ADM) decomposition of the classical spacetime and  present the Hamiltonian and the spatial and temporal diffeomorphism constraints first in terms of metric variables and then the Ashtekar-Barbero variables. We then specialize to the spatially-flat, homogeneous and isotropic spacetime to obtain the relevant Hamiltonian constraint which leads to the classical Friedmann and Raychaudhuri equations. Finally, we summarize the main results of the Hamiltonian formulation of anisotropic Bianchi-I spacetime and obtain the classical dynamical equations.

\subsection{The Hamiltonian formulation of general relativity in ADM decomposition and Ashtekar-Barbero variables}
\lb{sec:ADM}

The 3+1 foliation of globally hyperbolic spacetime was first applied to GR  by Arnowitt, Deser and Misner in their seminal work \cite{Arnowitt:1962hi}. %Later it is conventionally called ADM formalism in the literature which constitutes the basis for the Hamiltonian formulation of GR.
In this formalism, a 4-dimensional manifold $\mathcal M$ with a Lorentzian metric $g$ can be decomposed into $\mathcal M:=\Sigma\times \mathbb R$, where $\Sigma$ is the spacelike hypersurface and $ \mathbb R$ denotes the real line.  Correspondingly, the 4-metric of the manifold $g_{\mu\nu}$ with $\mu,\nu$ running from 0 to 4 can be expressed in terms of the lapse function  $N$,  the shift vector $N^a$ and the spatial metric $q_{ab}$ where $a,b=1,2,3$. To be specific, different components of the metric can be expressed as
\bq	
g_{00}=-N^2+N_aN^a,\quad g_{0a}=N_a,\quad g_{ab}=q_{ab}.
\eq
Using the above decomposition of the  4-metric in the  Einstein-Hilbert action of GR, one can obtain the canonical form of the action in the phase space spanned by the spatial metric and its conjugate momentum $\pi^{ab}$. This turns out to be  \cite{Arnowitt:1962hi}
\bq
\label{action1}
S=\int {\rm d}^4 x \left( \pi^{ab} \dot q_{ab} - N \mathcal H-N^a {\mathcal H}_a\right) .
\eq
Here the conjugate momentum $\pi^{ab}$ carries the information of the rate of change of the spatial metric and is given by
\bq
\pi^{ab}=\sqrt q\left(K^{ab}-Kq^{ab}\right) .
\eq
Here $K^{ab}$ stands for the extrinsic curvature of the hypersurface defined by $K_{ab}=(-\dot q_{ab}+D_aN_b+D_bN_a)/2N$ and   $q$ denotes the determinant of the spatial metric.  $D_a$ is the covariant derivative with respect to the spatial metric. The Hamiltonian constraint (or the scalar or the energy constraint) and the spatial diffeomorphism (or the momentum constraint) turn out to be respectively as  \cite{Arnowitt:1962hi}
\bqn
\lb{classical_constraints_ham}
{\mathcal H} &=&\frac{2 \kappa}{\sqrt{q}}\left(\pi^{ab}\pi_{ab}-\frac{\pi^2}{2}\right)-\frac{\sqrt q}{2\kappa}{}^{(3)}R\approx 0,\\
\lb{classical_constraints_diffeo}
{\mathcal H}_a &= & -2\partial_c\left(\gamma_{ab}\pi^{bc}\right)+\pi^{bc}\partial_a\gamma_{bc}\approx 0 ~,
\eqn
where $\kappa=8\pi G$ and ${}^{(3)}R$ denotes the intrinsic curvature of the 3-dimensional spatial hypersurface. Since  the lapse function and the shift vector in the  action (\ref{action1}) act as the Lagrange multipliers, the Hamiltonian and the diffeomorphism constraints weakly vanish on the  trajectories of the physical solutions as indicated by $``\approx0"$ in Eqs. (\ref{classical_constraints_ham})-(\ref{classical_constraints_diffeo}). These constraints are the first class constraints of the system and their Poisson brackets constitute the fundamental constraint algebra in GR. Recall that GR is a fully constrained system since the total Hamiltonian $H=\int \mathrm{d}^3 x\left(N \mathcal H+N^a {\mathcal H}_a\right) $ weakly vanishes on the dynamical trajectories of the physical solutions. As a result, the evolution of all the physical observables which are supposed to commute with all the first class constraints is frozen, leading to the problem of time. To extract physics from this frozen formalism we note that only those quantities are physically observable which are invariant under spacetime diffeomorphisms, such as Dirac observables which commute with the Hamiltonian.
To construct relevant Dirac observables which help extract dynamics we introduce appropriate reference fields or clock degrees of freedom  in the relational formalism \cite{komar,bergmann,  rovelli,dittrich1, dittrich2,  thiemann06,Thiemann:2004wk}. In the canonical formalism these reference fields have been studied in various contexts. In applications using framework of LQG, examples include the  reduced phase quantization which is based on quantizing phase space of gauge invariant quantities \cite{reduced1,Husain:2011tm,Giesel:2012rb,reduced2}, cosmological perturbations in presence of reference fields \cite{pert1,pert2,pert3,pert4,pert5}, and singularity resolution in LQC  \cite{Ashtekar:2006uz,Giesel:2020raf}.

Following Dirac's approach for quantization of constrained systems \cite{dirac}, and using the ADM formulation \cite{adm},
DeWitt obtained ``Einstein-Schr\"odinger equation" \cite{DeWitt:1967yk} which was later known as the Wheeler-DeWitt equation. In the case of finite degrees of freedom its implications were first studied in the mini-superspace setting by Misner \cite{misner}.
However, the convoluted form of the Hamiltonian which is non-polynomial in the metric variables makes it very difficult to obtain the physical solutions from the Wheeler-Dewitt equation in a general setting. Even in the mini-superspace setting of symmetry reduced models, such as a homogeneous and isotropic cosmological spacetime, where the relevant Wheeler-Dewitt equation can be exactly solved, physical solutions encounter singularities. An example is the case of the spatially-flat homogeneous and isotropic Friedmann-Lema\^itre-Robertson-Walker (FLRW) spacetime filled with a massless scalar field  \cite{Ashtekar:2006uz}. In fact, the consistent
quantum probability of singularity to occur in this model is unity including for wavefunctions which consist of arbitrary superpositions of expanding and contracting branches  \cite{Craig:2010ai,Craig:2010vf}. While it is possible to fine tune the avoidance of singularity in the WDW theory in some cases, such as in presence of suitable potentials or exotic matter such a resolution is often problematic and non-generic. This reveals one of the fundamental limitations of the WDW theory which not being based on a quantum theory of geometry fails to generically resolve the singularity. As we discuss later, this situation changes dramatically when using Ashtekar-Barbero variables in LQC.

In terms of the Ashtekar-Barbero variables -- the connection $A^i_a$ and the densitized triad $E^a_i$ \cite{Ashtekar:1986yd}, GR can be reformulated as a gauge field theory at least at a kinematical level using the internal SU(2) symmetry of variables.   They are defined explicitly by
\bq
E^a_i=  \sqrt{|q|}e^a_i~,\quad \quad A^i_a = \Gamma^i_a + \gamma K^i_a,
\eq
here $i,j$ are the internal SU(2) indices running from 1 to 3 and  $\gamma$ is the Barbero-Immirzi parameter, $e^a_i$ is the triad,  $K_a^i(= K_{ab}  e^{bi})$ is the extrinsic curvature with one of its indices projected onto the internal frame, and $\Gamma^i_a$ is the spin connection satisfying the condition that it be compatible with the triad, namely $D_ae^i_b=0$, resulting in
$\Gamma^i_a = -\frac{1}{2} \epsilon^{i j k} e_j^b\left(\partial_a e_b^k - \partial_b e_a^k + \delta_{m n} e_k^l e_a^m \partial_l e_b^n\right)$ with
$\epsilon^{i j k}$ being the Levi-Civita symbol. In terms of the new variables, the constraints of GR  can be rewritten as  \cite{Thiemann:2007pyv,BarroseSa:2000vx}
\bq
\label{constraints}
G_i=\partial_aE^a_i+\epsilon_{ijk}A^j_aE^{ak}\approx 0, ~~~~~~ \mathcal H_a=\frac{1}{\gamma}F^i_{ab}E^b_i-\frac{1+\gamma^2}{\gamma} K^i_aG_i\approx 0,
\eq
and
\bq
\label{hamclassical}
 \mathcal H=\frac{1}{\sqrt{|q|}}  E^a_j E^b_k \left(\epsilon_{i}^{j k} F_{ab}^i - \left(1 + \gamma^2\right) \left(K^j_a K_b^k - K^j_b K_a^k\right)\right) +\frac{1+\gamma^2}{\gamma}G^i\partial_a\frac{E^a_i}{\sqrt{|q|}}\approx 0~,
\eq
where the field strength of connection $A^i_a$ is given by $F^i_{ab}=\partial_a A^i_b-\partial_b A^i_a+\epsilon^i_{jk} A^j_aA^k_b$. The first constraint in Eq. (\ref{constraints}) is the Gauss constraint which generates the internal SU(2) symmetry. The Hamiltonian constraint  given in Eq. (\ref{hamclassical}) is a linear combination of a Euclidean term which is proportional to the field strength  and a Lorentzian term proportional to the extrinsic curvature and the Gauss constraint. In the quantum theory, the resulting form of the Hamiltonian constraint depends on the associated quantization ambiguities in loop quantizing these terms. In the case of the homogeneous models, due to the underlying symmetry reduction, it is possible to combine the Euclidean and Lorentzian terms \footnotetext{From a general perspective, in full LQG, it is possible to decompose the Lorentzian part of the Hamiltonian constraint into the one proportional to the Euclidean term and the spatial Ricci scalar. In a  spatially-flat FLRW universe, with a vanishing spatial curvature, the Lorentzian part becomes a multiple of the Euclidean term.}. Thus, a loop quantization based on utilizing this symmetry reduction is expected to be different from the one where quantization treats Euclidean and Lorentzian terms independently. The standard LQC is based on the former strategy while the latter method results in modified versions of LQC.
We would later see that different treatments of the Euclidean and Lorentzian parts result in different physical predictions than from the standard LQC.

\subsection{Spatially-flat, homogeneous and isotropic spacetime: classical aspects}
\lb{sec:spatially_flat_FLRW}

Given that most of the analysis in LQC so far has focused on understanding implications for the spatially-flat, homogeneous, isotropic model, in the following we provide a brief overview of the Hamiltonian aspects in the classical theory for this model.
%The main arena for exploring the quantum gravity effects in the current cosmic microwave background (CMB) observations is
The spacetime metric in the spatially-flat  FLRW universe is given by
\bq
\ds s^2=-N^2\ds t^2+a^2(t)\delta_{ab}\ds x^a \ds x^b,
\eq
where $a(t)$ denotes the scale factor of the universe. Its spatial topology can be either a 3-torus $\mathbb T^3$ or non-compact $\mathbb R^3$.  For the torus  $\mathbb T^3$, the co-moving length of each spatial direction is naturally confined within the range $x_a\in(0,l_o)$ which makes the spatial integrals in the Hamiltonian finite. For the $\mathbb R^3$ topology, a fiducial cell ${\cal V}$ with co-moving volume $ V_o$ is required. All the spatial integrals are restricted to the fiducial cell so that no divergences would be encountered. In this way, there is also a well-defined symplectic structure even with $\mathbb R^3$ spatial topology. In the following analysis, we take the $\mathbb R^3$ spatial topology.  In addition to the gravitational degrees of freedom,  for the matter content, we consider the simplest case of  a massless scalar field $\phi$ which serves as a reference field for relational dynamics. Thus, the mini-superspace of the spatially-flat FLRW universe consists of the scale factor $a$ and its conjugate momentum $\pi_a$ as well as the canonical pair  $\phi$ and $\pi_\phi$. In terms of these two canonical pairs, the action in (\ref{action1}) with additional contributions from the matter sector reduces to
\bq
S=\int \ds t \Big\{ V_o \left(\pi_\phi \dot{ \phi}+\pi_a\dot a\right)-N\mathcal H\Big\},
\eq
which implies the standard Poisson brackets  $\{a, \pi_a\}=\{\phi,\pi_\phi\}=1/V_o$.  Note that due to the homogeneity of the background spacetime, the spatial diffeomorphism constraint identically vanishes while the background Hamiltonian constraint  reduces to
\bq
\mathcal H=V_o\left(-\frac{\kappa \pi^2_a}{12 a}+\frac{\pi^2_\phi}{2a^3}\right).
\eq
The corresponding Hamilton's equations can be obtained by evaluating  the Poisson bracket between the canonical variables and the background Hamiltonian constraint. To be specific,  one finds
\bq
\dot a =\{a, \mathcal H\}=- \frac{\kappa \pi_a}{6a},\quad \quad \dot \pi_a=\frac{3\pi^2_\phi}{2a^4}-\frac{\kappa\pi^2_a}{12a^2}.
\eq
It is straightforward to obtain the Friedmann and Raychaudhuri equations from above equations.
The classical Friedmann equation can be obtained by squaring the Hamilton's equation of the scale factor and then expressing $\pi^2_a$ in terms of the energy density of the scalar field resulting from the vanishing of the background Hamiltonian constraint. On the other hand, the Raychaudhuri equation can be derived from the Hamilton's equations of the scale factor and its conjugate momentum. Collectively, the classical Friedmann and Raychaudhuri  equations take the form
\bq
H^2=\frac{8\pi G}{3}\rho, \quad \quad \frac{\ddot a}{a}=-4\pi G\left(\rho+3P\right),
\eq
where the Hubble rate is defined via $H=\dot a/a$ and for a massless scalar field $P=\rho=\dot \phi^2/2$.
Although both of the action and the background Hamiltonian contain the fiducial volume $V_o$, the resulting equations of motion are independent of $V_o$, implying that the classical dynamics remains unaffected by any other  choice of the fiducial cell.
Using the Hamilton's equations for the matter sector, it is also easy to check that the scalar field satisfies the Klein-Gordon equation or equivalently the matter-energy conservation law:
\bq
\dot \rho + 3 H (\rho + P) = 0.
\eq
We can easily see these dynamical equations result in a singularity for a given choice of matter.
For matter which has a fixed equation of state
$w = P/\rho$, the conservation law results in $\rho \propto a^{-3(1+w)}$. For matter satisfying weak energy condition, $\rho \geq 0$ and $\rho + P \geq 0$,  as the scale factor approaches zero the energy density and pressure diverge. This corresponds to the big bang singularity in the backward evolution (or the big crunch singularity in the forward evolution) where the Friedmann and Raychaudhuri equations break down.

Let us now discuss how to obtain these equations using the Ashtekar-Barbero variables.
Owing to the homogeneity of the spacetime, the connection $A^i_a$ and the triad $E^a_i$ can be expressed in terms of their isotropic counterparts $c$ and $p$ as \cite{Ashtekar:2003hd}
\bq
\lb{spatially_flat_Ashtekar_variables}
A^i_a=cV^{-1/3}_o\mathring{\omega}^i_a,\quad E^a_i=p V^{-2/3}_o\sqrt{\mathring q}\mathring{e}^a_i~.
\eq
Here $\mathring{e}^a_i$ and $\mathring{\omega}^i_a$ are the fiducial triads and co-triads compatible with the fiducial metric which is simply $\mathring q_{ab}=\delta_{ab}$ in the spatially-flat FLRW spacetime. The symmetry reduced isotropic connection and triad satisfy the Poisson bracket $\{c,p\}=8\pi G\gamma/3$. The triad is kinematically related to the metric variables as $|p|=V^{2/3}_oa^2$ where the modulus sign over the triad arises to account  two possible orientations. The relationship between the connection and the scale factor is a dynamical one to be obtained from the Hamilton's equation for the triad. In the classical theory it is given by $c=\gamma V^{1/3}_o\dot a$. It is important to note that the latter relation changes in LQC.

As discussed later, in improved dynamics of LQC \cite{Ashtekar:2006wn}, it is more convenient to work with another set of phase space variables related to $(c,p)$ variables as \cite{Ashtekar:2007em}
\bq\label{eq:bv-variables}
b:=c/|p|^{1/2}, ~~~ v:=sgn(p)|p|^{\frac{3}{2}},
\eq
where $sgn(p)=\pm 1$ for the same/opposite orientation of the physical and fiducial triads. These variables satisfy the Poisson bracket $\{b,v\}=4\pi G \gamma$.
In term of the new variables $b$ and $v$, the background Hamiltonian for the spatially-flat FLRW universe can be written in the form
\bq
\lb{classical_FLRW_Hamiltonian_constraint}
\mathcal H=-\frac{3b^2|v|}{8\pi G\gamma^2}+\frac{p^2_\phi}{2|v|},
\eq
with $p_\phi=V_o \pi_\phi$.
It is straightforward to derive the Hamilton's equations for $v$ and $b$ which result in the Friedmann and Raychaudhuri equations using the vanishing of the background Hamiltonian constraint.
From the Hamilton's equations of $v$ and $\phi$, it is straightforward to obtain the relation
\bq
\frac{\ds v}{\ds \phi}=\frac{3b|v|^2}{p_\phi \gamma}.
\eq
Let us note that in the classical theory $b$ is related with the Hubble rate via $b=\gamma H$, then using the classical Friedmann equation, the above differential equation turns out to be equivalent to
$\ds \phi=\frac{sgn(p_\phi)}{\sqrt{12\pi G}|v|}\ds v$, leading to the generic solution
\bq
\phi=\pm\frac{1}{\sqrt{12\pi G}}\ln|\frac{v}{v_0}|+\phi_0,
\eq
where $v_0$ and $\phi_0$ are integration constants and $\pm$ sign is determined by the sign of the constant momentum. The above solutions represent two disjoint trajectories  which describe an expanding and a contracting universe respectively. The singularity is inevitable when the volume reaches zero in the past of an expanding universe which thus emerges from a big bang singularity or in the future of a contracting universe that ends up with a big crunch singularity. A plot of these solutions is shown in Fig. 1 as dashed curves.

\subsection{The Hamiltonian formulation of anisotropic spacetime: Bianchi-I model}
\lb{Hamiltonian_Bianchi-I}

The Bianchi-I spacetime is one of the simplest models in which the effects of anisotropies on the physics near the classical singularities can be systematically investigated. It has vanishing intrinsic curvature and its isotropic limit is the spatially-flat FLRW universe. Unlike in the isotropic models where the big bang singularity is characterized by a vanishing scale factor of the universe, the structure of the singularities in the anisotropic spacetimes becomes much richer since the directional scale factors can evolve in different ways when the singularity is approached. As discussed later in the next section, the shape of the singularity can be of a cigar, a pancake, a barrel or a point depending on the way directional scale factors approach zero.
As one of the simplest anisotropic models to study, the dynamics of Bianchi-I spacetime is important to understand as it plays a vital role in  understanding the approach to singularities in the classical Bianchi-IX spacetime and in particular the Mixmaster dynamics  \cite{Berger:2002st}.

In the following,  we consider  the simplest case of a homogeneous Bianchi-I spacetime with a manifold $\mathcal M:=\Sigma\times \mathbb R$, where $\Sigma$ is topologically flat, and its metric is given by
\bq
\ds s^2=-N^2 \ds t^2+a^2_1(t)\ds x^2+a^2_2(t)\ds y^2+a^2_3(t)\ds z^2,
\eq
where $a_i$ with $i=1,2,3$ are the directional scale factors.
Similar to the case of the spatially-flat FLRW spacetime, in order to have a well-defined symplectic structure, a fiducial cell with $l_i$ denoting its coordinate length along each side needs to be introduced. Therefore, the fiducial volume equals $V_o=l_1l_2l_3$. Due to the homogeneity of the spacetime, the Ashtekar-Barbero variables, namely the connection $A^i_a$ and the densitized triad $E^a_i$ can be expressed in terms of $c_i$ and $p_i$ along each direction as \cite{Ashtekar:2009vc}
\bq
A^i_a=\frac{c_i}{l_i}\mathring{\omega}^i_a,\quad E^a_i=p_i l_iV^{-1}_o\sqrt{\mathring q}\mathring{e}^a_i~.
\eq
Note there is no summation over index `i' in the above definitions.
These canonical phase space pairs satisfy $\{c_i,p_j\}=8\pi G \gamma \delta_{ij}$. Moreover, the triads are related to the metric variables by
\bq
p_1=\epsilon_1 l_2l_3a_2a_3,\quad  p_2=\epsilon_2 l_1l_3a_1a_3,\quad p_3=\epsilon_3 l_1l_2a_1a_2,
\eq
with $\epsilon_i=\pm 1$ determined by  the triad orientation.  Since we are not considering fermions in the matter content, without any loss of generality, we choose the plus sign for the triads. The relationship between the connection and the metric variables is determined from the Hamilton's equations. Due to the homogeneity of the spacetime, the only non-vanishing constraint is the Hamiltonian constraint which  in term of the connection and triad takes the form \cite{Ashtekar:2009vc}
\bq
\mathcal H=-\frac{N}{8\pi G \gamma^2 v}\left(c_1p_1c_2p_2+c_1p_1c_3p_3+c_2p_2c_3p_3\right)+Nv\rho,
\eq
where $v=V_oa_1a_2a_3$ denotes the physical volume of the fiducial cell and $\rho$ is the matter energy density. From the Hamilton's equations of the triads one finds
\bq
c_i=\gamma l_i \dot a_i,
\eq
where the lapse is already set to unity so that the time derivative is with respect to the cosmic time. Using the above relation in the Hamiltonian constraint, it is easily seen that the energy density of the matter sector is related with the directional Hubble rate via
\bq
\kappa \rho=H_1H_2+H_2H_3+H_3H_1.
\eq
Now defining the mean Hubble rate $H=\left(H_1+H_2+H_3\right)/3$, we can obtain a generalized Friedmann equation in terms of the mean Hubble rate which takes the form
\bq
H^2=\frac{\dot a ^2}{a^2}=\frac{\kappa}{3}\rho+\frac{\Sigma^2}{a^6},
\eq
where the mean scale factor is defined by  $a=(a_1a_2a_3)^{1/3}$ and $\Sigma^2$ is related with the shear scalar $\sigma^2$ and directional Hubble rates $H_i$ via
\bq
\Sigma^2=\frac{1}{6}\sigma^2a^6=\frac{a^6}{18}\left(\left(H_1-H_2\right)^2+\left(H_2-H_3\right)^2+\left(H_3-H_1\right)^2\right).
\eq
Note that the above relation is kinematical and holds as it is in LQC.
For the matter content which has no anisotropic stress, such as a massless/massive scalar field which is minimally coupled to gravity, $\Sigma^2$ is a constant of motion \cite{Chiou:2007sp}. As a result, the anisotropies behave like a perfect fluid with an effective equation of state $w=1$ and   the shear scalar diverges as $\sigma^2\propto a^{-6}$ which is faster than any perfect fluid with an equation of state $w<1$ when the singularity at vanishing mean scale factor is approached.  In those cases, the anisotropies dominate near the classical singularities. Finally, for the matter content which is composed of a massive scalar field, with its matter Hamiltonian given by the familiar form $\mathcal H_m=p^2_\phi/2v+vU(\phi)$, from the Hamilton's equations of the scalar field and its momentum, one can obtain the standard Klein-Gordon equation
\bq
\ddot \phi+3H\dot \phi+U_{,\phi}=0,
\eq
where $H$ is the mean Hubble rate and $U_{,\phi}$ denotes the derivative of the potential with respect to the scalar field. An important implication from the Klein-Gordon equation is that the dynamics of the scalar field is determined by the averaged effects of the anisotropies, namely the mean Hubble rate and the scalar field itself can not detect the fine structure of the anisotropies.

\section{Nature of classical singularities: types, strength and shapes}
\lb{Sec:nature_of_singularities}
\renewcommand{\theequation}{3.\arabic{equation}}\setcounter{equation}{0}

So far we have discussed the Hamiltonian formulation of isotropic and Bianchi-I model using the metric as well as Ashtekar-Barbero variables and obtained the dynamical equations which signal existence of singularities.
In this section, we overview some of the properties  of the singularities encountered in this cosmological setting. According to singularity theorems of Penrose, Hawking and Geroch \cite{Geroch:1968ut,Hawking:1970zqf}, singularities are the boundaries of the spacetime beyond which the null/timelike geodesics can not be extended in a unique deterministic manner. However, an important and complimentary aspect of singularities is their physical nature in terms of their strength which in literature has played a useful role in gaining insights on the nature of shell crossing and naked singularities in GR.
%In the cosmological context, geodesic incompleteness and singularities being strong are synonymous.
%They are usually regarded as the  embodiment of the limitations of the classical theory  and  are supposed to be resolved in the quantum theory of gravity.
In the following, we will first briefly discuss different types  and strength of the singularities  in the isotropic spacetimes. In the second part of the section, taking Bianchi-I model as an example, we address the different shapes of the singularities in an anisotropic spacetime.

\subsection{Types and strength of the singularities}
In the classical cosmology, as long as the null energy condition $(\rho + P \geq 0)$ is satisfied, the spatially-flat isotropic FLRW universe would inevitably encounter the big bang/crunch singularity when evolved backwards in time. Not even the inflationary scenario can help evade the big bang singularity as the Borde-Guth-Vilenkin theorem proves \cite{Borde:2001nh}.  In addition to the big bang/crunch singularity, there can be other types of cosmological singularities, depending on the behavior of the scale factor $a$, the  energy density $\rho$ and  the pressure $P$ as the singularity is approached \cite{Cattoen:2005dx,Fernandez-Jambrina:2006tkb}. The latter two observables determine the behavior of the Ricci scalar $R$ in a spatially-flat universe since
\bq
R=8\pi G\left(\rho-3P\right).
\eq
In general, for the matter with a non-dissipative equation of state  $P=P(\rho)$, all the known cosmological singularities can be classified into the following categories \cite{Nojiri:2005sx}:

\begin{itemize}

\item \textit{Big Bang}/\textit{Crunch} singularity: This type of singularities is characterized by a vanishing volume (scale factor) of universe with the infinite energy density and pressure, implying that the Ricci scalar also blows up when the singularity is approached. The null energy condition is satisfied at these singularities. They signal the ultimate future/beginning of  a contracting/expanding classical universe which can be reached at a finite proper time along the geodesics.
\\
\item \textit{Big Rip} or  \textit{Type-I} singularity:  At this type of singularity, all of the relevant quantities, such as the scalar factor, the energy density, the pressure as well as the Ricci curvature, become divergent. The null energy condition is also violated at the singularity. An example of this singularity can be found in the models with a phantom field of the equation of state $w<-1$. If we consider a phantom fluid with a fixed equation of state, then the conservation law results in $\rho \propto a^{-3(1+w)}$ which shows that the energy density blows  up as the scale factor becomes divergent. The pressure and the Ricci scalar have the same fate.
\\
\item  \textit{Sudden} or \textit{Type-II} singularity: This singularity is featured by a finite scale factor and the energy density. However, both the pressure and the Ricci scalar become infinite when the singularity is reached  \cite{Barrow:2004hk}. Therefore, the dominant energy conditions is violated for this type of singularities.
\\
\item \textit{Big Freeze} or \textit{Type-III} singularity: This type of singularity also occurs at finite value of the scale factor but all the other physical observables, such as the energy density, the pressure and the Ricci scalar become divergent \cite{Bouhmadi-Lopez:2006fwq}.
\\
\item \textit{Type-IV} singularity: As compared with all of its cousins discussed above, this type of singularities behaves normally when assessed by the behavior of the scale factor, the energy density, the pressure and the Ricci scalar since all of these quantities are finite valued. The divergence shows up when the curvature derivatives are computed \cite{Barrow:2004he}.
\\
\end{itemize}
The analysis of the energy conditions for the last three types of the singularities are model dependent \cite{Cattoen:2005dx}. Although the singularities discussed above are related with the divergence of physical observables or their derivatives, not all of these singularities can be regarded as the physical ones.  Depending on the strength of the tidal forces at the singularities, the singularities can be either  the strong singularities or the  weak singularities \cite{Ellis:1977pj,Tipler:1977zza,Krolak:1986}. Intuitively speaking,  at the strong singularities, the tidal forces are  strong enough to destroy any objects or detectors that try to pass through while the weak singularities still allow the passage of strong detectors. The strength of the singularities can be studied in a quantitative way using the projection of the Ricci tensor along the tangent direction of the geodesics, namely,
\bq
\lb{strong_singularity_criterion}
\int^{\tau}_0\mathrm{d}\tau R_{ab} u^au^b.
\eq
If the above integral becomes divergent when $\tau$ asymptotes to the finite proper time at which the  singularity occurs, then this singularity is regarded as strong by Kr\'olak \cite{Krolak:1986}. If the above integral remains finite then the singularity is weak.  A more restrictive version of the above criterion is to compute the double integral of the same integrand as considered by Tipler \cite{Tipler:1977zza}. Straightforward calculations of the integral (\ref{strong_singularity_criterion}) in the FLRW universe reveals that the result of the integral is finite when the scale factor is a finite value if only  its second- or higher-order derivatives are divergent. As a result, the big bang/crunch/rip singularity is the strong singularity while the Type-II/IV singularities are weak. For the Type-III singularity, it is strong according to Kr\'olak's condition and weak by Tipler's condition. It is believed that strong singularities are also the ones where geodesics can not be extended.

\subsection{Shapes of the singularities}

In the isotropic models, the scale factor in each direction behaves in a uniform way and  the singularities that are reached by a vanishing scale factor take the form of a point and thus are called the point singularities. In contrast, the structure of the singularities becomes much richer in presence of anisotropies where the directional scale factors can approach the singularities in different manners.  Depending on the behavior of the directional scale factor when a singularity is approached, there can be different shapes as listed below \cite{Thorne:1967zz}:

\begin{itemize}

\item  \textit{Point} singularity: The directional scale factors simultaneously reach zero at the singularities, mimicking the behavior of the point-like singularity in the isotropic model. For this type of singularities to occur, the matter energy density must dominate over that of the anisotropic shear which implies that the equation of state of the matter must satisfy $w\ge 1$. An example in this case is an anisotropic universe dominated by  a single scalar field with a negative potential.
\\
\item  \textit{Barrel} singularity: This singularity appears when one of the directional scale factors tends to a non-zero constant value while the other two  become vanishing.
\\
\item  \textit{Pancake} singularity: In contrast to the Barrel singularity, the Pancake singularity is characterized by the vanishing of one of the directional scale factors  with the other two approaching non-zero constants.
\\
\item  \textit{Cigar} singularity: This singularity emerges when one of the directional scale factors becomes divergent while the other two tend to vanish.
\\
\end{itemize}

In presence of anisotropies, a point like singularity can occur only when matter energy density dominates evolution over the anisotropic shear. Since the anisotropic shear behaves as $\sigma^2 \propto a^{-6}$ this means that for all matter with equation of state $w < 1$, the shape of the singularity is not point like. In general, the singularity is a cigar like singularity. In particular,  when the matter is absent,  only the pancake and the cigar singularities can occur in the  Bianchi-I spacetime. This can be easily understood by introducing the Kasner exponents $k_c$ with $c=1,2,3$ in which the directional scale factors of the classical solutions scale as $a_c\propto t^{k_c}$. For the vacuum solution, there are two constraints on the Kasner exponents, namely  $k_1+k_2+k_3=1$ and $k^2_1+k^2_2+k^2_3=1$. As a result, point singularity is impossible to occur since it requires $k_1=k_2=k_3=1/3$ which violates the second condition.  For the same reason, neither the barrel singularity which requires $k_1=0,k_2=k_3=1/2$ is possible. In contrast, the pancake singularity can be reached by $k_1=k_2=0, k_3=1$ while the cigar singularity can be realized by choosing one of the Kasner exponents to be a small negative number and then solve for the other two. Thus, the allowed shapes of the singularity is essentially determined by the initial conditions on the matter content and anisotropies. When the Bianchi-I universe is filled with dust, only the cigar and pancake singularities  are  possible to occur. In  contrast, if  the matter content is composed of stiff matter, then the barrel, cigar and point  singularities can form. \\

Having discussed the nature of classical singularities in some detail, it is pertinent to ask whether quantum geometry effects as understood in LQC resolve all different types of singularities, and what happens to the shape of the bounce. To answer these questions we need to go beyond the classical setting and perform a loop quantization of the isotropic and anisotropic models.

\section{Loop quantum cosmology: spatially-flat isotropic model}
\lb{sec:isotropic_model}
\renewcommand{\theequation}{4.\arabic{equation}}\setcounter{equation}{0}

In this section, we overview the key steps of the loop quantization of a spatially-flat FLRW universe as originally proposed in \cite{Ashtekar:2006rx,Ashtekar:2006uz,Ashtekar:2006wn}.  It provides a framework where all the aspects concerning the construction of a quantum theory of gravity, such as the quantum Hamiltonian constraint, the physical Hilbert space, the inner product, the Dirac observables and the semi-classical states, are obtained systematically.  The construction of this model also serves as a prototype for applying the techniques of LQG to symmetry reduced spacetimes with the purpose of understanding the singularity resolution and the effects of quantum gravity therein. Later, it was succeeded by a series of remarkable works exploring the quantum gravity effects in the spacetimes with more complicated structure, such as those with a cosmological constant \cite{Bentivegna:2008bg,Kaminski:2009pc,Pawlowski:2011zf}, radiation \cite{Pawlowski:2014fba}, the spatial curvature \cite{Ashtekar:2006es,Szulc:2006ep,Vandersloot:2006ws,Szulc:2007uk}, the anisotropies \cite{Chiou:2007sp,Ashtekar:2009vc, Ashtekar:2009um,Wilson-Ewing:2010lkm,Chiou:2006qq, Martin-Benito:2008dfr, Martin-Benito:2009xaf}, in presence of inhomogeneities \cite{Martin-Benito:2008eza,Brizuela:2009nk,Garay:2010sk,Martin-Benito:2010vep,Martin-Benito:2010dge}, as well as other variants resulting from different quantization prescriptions \cite{Yang:2009fp}. Although  the loop quantization of  these different cosmological  models exhibits  their own  properties in the Planck regime, there also exists one common feature originating from introducing the minimal area gap in LQG in the course of  quantization,  that is the generic  resolution of the spacetime curvature singularities \cite{Singh:2009mz,Singh:2011gp,Singh:2014fsy,Saini:2017ipg,Saini:2017ggt,Saini:2018tto,Singh:2010qa}. In the following, we start with the main steps of the loop quantization of the spatially-flat FLRW universe. A key step in this quantization procedure deals with a careful assignment of the area of the loop over which holonomies are considered. In the early stage of LQC this area was taken to be the coordinate area which resulted
in a quantum theory that suffers from several drawbacks, for example the dependence of the bounce density on the initial conditions and the fiducial cell and lack of infra-red limit as GR when matter violates strong energy condition \cite{Corichi:2008zb}.  It turns out that the quantization emerging from considering physical areas, also known as improved dynamics or $\bar \mu$ scheme in LQC  \cite{Ashtekar:2006rx}, is the only physically viable choice in the isotropic models \cite{Corichi:2008zb,Corichi:2009pp}. In the  improved dynamics of LQC it is more convenient to work with $b$ and $v$ variables obtained from the $c$ and $p$ variables using a canonical transformation (see \eqref{eq:bv-variables}) \cite{Ashtekar:2007em}. In the following, we first discuss the $\bar \mu$ scheme in LQC using volume representation and then comment on the uniqueness of the $\bar \mu$ scheme.

\subsection{Loop quantization of the spatially-flat FLRW universe}
\lb{sec:LQC_in_spatially_flat_universe}

In this subsection, applying the techniques  in full LQG,  we outline the construction of the Hamiltonian constraint operator for the spatially-flat FLRW universe in LQC as originally developed in \cite{Ashtekar:2006uz,Ashtekar:2006wn}.  The fundamental variables for loop quantization in LQG is the holonomies of the connection along a path $l$ and the fluxes of the triads over a  2-surface $S$, to be specific, they are formally given by
 \bq
 h_l=\mathcal P\mathrm{exp}\left(\int_l~ \ds l ~A^i_a\tau_i \frac{\ds x^a}{\ds l} \right),\quad \quad E_i=\int_S n_a E^a_i\mathrm{d}^2s,
 \eq
 where the path is parameterized by  $x^a(l)$, $\mathcal P$ stands for the path-ordered product, $n_a$ is the unit normal to the surface $S$ and $\tau_i=-i\sigma_i/2$ with $\sigma_i$ being the Pauli matrices. In the spatially-flat FLRW universe, due  to the homogeneity and isotropy of the spacetime, the integrals in the above definitions can be computed analytically. For the holonomies, we consider a path along the edge $\mu\mathring{e}^a_i$, yielding
\bqn
h^{(\mu)}_i&=&\exp \int^{ V^{1/3}_0}_0 \tau_j A^j_a~\mu\mathring{e}^a_i ~ \mathrm{d}l=\cos(\frac{\mu c}{2})\mathbb{I}+2\tau_i \sin(\frac{\mu c }{2}),
\eqn
where the symmetry reduced connection  (\ref{spatially_flat_Ashtekar_variables}) is used. Similarly, one can explicitly work out the fluxes of the triads which turn out to be proportional to the symmetry reduced triad $p$ \cite{Ashtekar:2003hd}. Therefore, the elementary classical variables of the geometric sector for loop quantization of  a spatially-flat FLRW universe are the almost periodic functions of the connection $N_\mu=e^{i\mu c/2}$  and  the isotropic triad $p$. They form the so-called abstract $\star$-algebra which has a unique representation in the kinematical Hilbert space $\mathcal H_\mathrm{kin}$: $L^2(\mathbb{R}_\mathrm{Bohr},\mathrm{d}\mu_\mathrm{Bohr})$ \cite{Ashtekar:2012cm,Engle:2016hei,Engle:2016zac}.  $\mathcal H_\mathrm{kin}$ consists of the square integrable functions on the Bohr compactification of the real line with the basis states given by  $N_\mu:=\langle c|\mu\rangle$. The label of the states $\mu$ is essentially discrete as can be seen from the inner product of two basis states
\bq
\langle \mu_1|\mu_2\rangle=\lim_{D\rightarrow\infty}\frac{1}{2D}\int^D_{-D}\mathrm{d}c\langle \mu_1|c\rangle\langle c|\mu_2\rangle=\delta_{\mu_1,\mu_2},
\eq
where $\delta_{\mu_1,\mu_2}$ is the Kronecker delta.  A general state in $\mathcal H_\mathrm{kin}$ is a countable sum of the basis states as $|\Psi\rangle=\sum_n \alpha_n|\mu_n\rangle$ with the complex coefficients satisfying the condition $\sum_n |\alpha_n|^2<\infty$. There are unambiguous operator representations  of $N_\mu$ and the triad $p$ which act on the basis states in the way \cite{Ashtekar:2006uz}
\bq
\widehat N_{\mu_0} |\mu\rangle=|\mu+\mu_0\rangle,\quad  \hat p|\mu\rangle=\frac{8\pi \gamma l^2_\mathrm{pl}}{6}\mu|\mu\rangle,
\eq
here $\mu_0$ is a constant and the Planck length $l^2_\mathrm{pl}=G \hbar$.  From the operator  $\widehat N_{\mu} $, one can easily find the action of the holonomy operator $\hat h^{(\mu)}_i$ on the states in $\mathcal H_\mathrm{kin}$ \cite{Ashtekar:2006wn}. The operators $\hat h^{(\mu)}_i$ and $\hat p$ are the elementary building blocks for constructing the quantum Hamiltonian constraint in LQC.
Now it is time to take the advantage of the symmetry of the spacetime under consideration, due to homogeneity and isotropy, it is obvious that the Gauss  and the momentum constraints given in (\ref{constraints})  vanish identically. Therefore, only the Hamiltonian constraint (\ref{hamclassical}) remains to be implemented. Furthermore, the Lorentzian term in the Hamiltonian constraint (\ref{hamclassical})  turns out to be a multiple of the Euclidean term in the spatially-flat FLRW universe. As a result, in standard LQC, the geometrical sector of the quantum Hamiltonian constraint is constructed from its classical counterpart
\bq
\lb{geo_sector}
\mathcal H_\mathrm{g}=-\int_\mathcal{V}\mathrm{d}^3x \frac{N}{2\kappa\gamma^2\sqrt{|q|}}  E^a_j E^b_k \epsilon_{i}^{j k} F_{ab}^i,
\eq
which absorbs the  Lorentzian term into the coefficient of the  Euclidean term. A separate treatment of the  Lorentzian term can result in phenomenologically distinct models  \cite{Yang:2009fp,Li:2018opr} as will be discussed in Sec. 7. Next we need to express the above Hamiltonian in terms of the holonomies and the triad. It contains two pieces, the first piece can be written as \cite{Thiemann:2007pyv,Thiemann:1996aw}
\bq
\lb{Thiemann_identity}
\frac{\epsilon_{i j k}}{\sqrt{|q|}}E^{aj}E^{bk}=\sum_{k}\frac{sgn(p)}{2\pi \gamma G\mu V^{1/3}_o }\mathring{\epsilon}^{abc}\mathring{\omega}^k_c\mathrm{Tr}\left(h^{(\mu)}_k\Big\{\left(h^{(\mu)}_k\right)^{-1},V\Big\}\tau_i\right),
\eq
where $V=|p|^{3/2}$ is the physical volume and `Tr' means to take the trace of the parenthesis. The other piece is the field strength $F^i_{ab}$ which can be expressed in terms of the holonomies over a square in the $j$-$k$ plane spanned by the triads $\mathring{e}^a_j$ and  $\mathring{e}^b_k$, namely,
\bq
\lb{field_strength}
F^i_{ab}=-2 \lim_{\mu\rightarrow 0}\frac{\mathrm{Tr}\Big[ \tau^i\left(h^{(\mu)}_{\square_{jk}}-\mathbb{I}\right)\Big]}{\mu^2 V^{2/3}_0}  ~\mathring{\omega}^j_a~{}\mathring{\omega}^k_b,
\eq
where $h^{(\mu)}_{\square_{jk}}=h^{(\mu)}_jh^{(\mu)}_k(h^{(\mu)}_j)^{-1}(h^{(\mu)}_k)^{-1}$ is the holonomy over a square loop. In the classical theory, the field strength is a local quantity as the area of the square vanishes when $\mu\rightarrow 0$. When shifting to the quantum theory, the holonomies  and the volume in (\ref{Thiemann_identity})-(\ref{field_strength}) should be promoted to their operator analogs  and the Poisson brackets be replaced by the commutators. In addition to these standard operations,  extra care should be taken of the area of the square used to define the field strength. The key point is that in LQG, the area operator has a discrete spectrum with a minimal non-zero eigenvalue. This feature is inherited in LQC in the sense that a non-local field strength operator  is defined  by shrinking the area of the square to the minimal area gap available for the homogeneous spacetime  which is usually denoted by  $\Delta=4\sqrt 3 \pi \gamma l^2_\mathrm{pl}$. Relating the physical area of the square $(a^2\mu^2V^{2/3}_o)$ with this minimal area gap, we can fix the value of $\mu$ to be
\bq
\lb{mu_bar_scheme}
\mu=\bar \mu:=\frac{\lambda}{\sqrt{|p|}},
\eq
with $\lambda=\sqrt \Delta$. This choice of $\mu$ leads to the well-known $\bar \mu$ scheme in LQC. Since $\bar \mu$ explicitly depends on the triad, $\widehat N_{\bar \mu}$ no longer acts as a shift operator on the eigenstates of the triad operator $\hat p$. In order to make the quantum Hamiltonian constraint operator a difference operator with uniform step size,  it is  convenient to work with the eigenstates of the volume operator on which $\widehat N_{\bar \mu}$ behaves like a shift operator again. Specifically, we have  \cite{Ashtekar:2006wn, Vandersloot:2006gga}
\bq
\hat V|v\rangle=\left(\frac{8\pi \gamma}{6}\right)^{3/2}\frac{|v|}{K}l^3_\mathrm{pl}|v\rangle,\quad \quad \widehat N_{\bar \mu}|v\rangle=|v+1\rangle,
\eq
where $K=\frac{2}{3\sqrt{3\sqrt 3}}$ and two sets of basis states, i.e. $|v\rangle$ and $|\mu\rangle$ are related with each other via $v=K \mathrm{sgn}(\mu) |\mu|^{3/2}$. To extract dynamics, we consider a massless scalar field which serves as a matter clock, whose classical and Fock quantized Hamiltonian is given by
\bq
\lb{matter_sector}
\mathcal H_\mathrm{m}=\frac{p^2_\phi}{2|p|^{3/2}}\rightarrow \widehat{\mathcal H}_\mathrm{m}=-\frac{\widehat{|p|^{-3/2}}\partial^2}{2\partial\phi^2} .
\eq
Note that unlike the gravitational sector,  the Schr\"odinger representation is employed for the scalar field. Now according to the Dirac's quantization approach for the constrained system, the physical state $\Psi(v,\phi):=\langle \Psi|v\rangle$ must be annihilated by the quantum Hamiltonian constraint operator, that is
\bq
\lb{Diracs_prescription}
\widehat{\mathcal H}_0  \Psi(v,\phi)=\left(\widehat{\mathcal H}_\mathrm{g}+\widehat{\mathcal H}_\mathrm{m}\right) \Psi(v,\phi)=0.
\eq
Combining (\ref{geo_sector}), (\ref{Thiemann_identity}), (\ref{field_strength}), (\ref{matter_sector}) and (\ref{Diracs_prescription}), we can obtain a quantum difference equation which governs the dynamical evolution of the physical state and reads explicitly \cite{Ashtekar:2006wn}
\bq
\lb{quantum_difference_equation}
\partial^2_\phi \Psi(v,\phi)=[B(v)]^{-1}\left(C^+\Psi(v+4,\phi)+C^0\Psi(v,\phi)+C^-\Psi(v-4,\phi)\right)=:-\Theta\Psi(v,\phi),
\eq
where the volume-dependent coefficients are given by
\bqn
B(v)&=&\left(\frac{3}{2}\right)^3K|v|||v+1|^{1/3}-|v-1|^{1/3}|^3,~ C^+(v)=\frac{3\pi K G}{8}|v+2|||v+1|-|v+3||,\nb\\
C^-(v)&=&C^+(v-4),\quad \quad C^0(v)=-C^+(v)-C^-(v).
\eqn
\begin{figure}
{
\hspace*{-1cm}
\includegraphics[width=12cm]{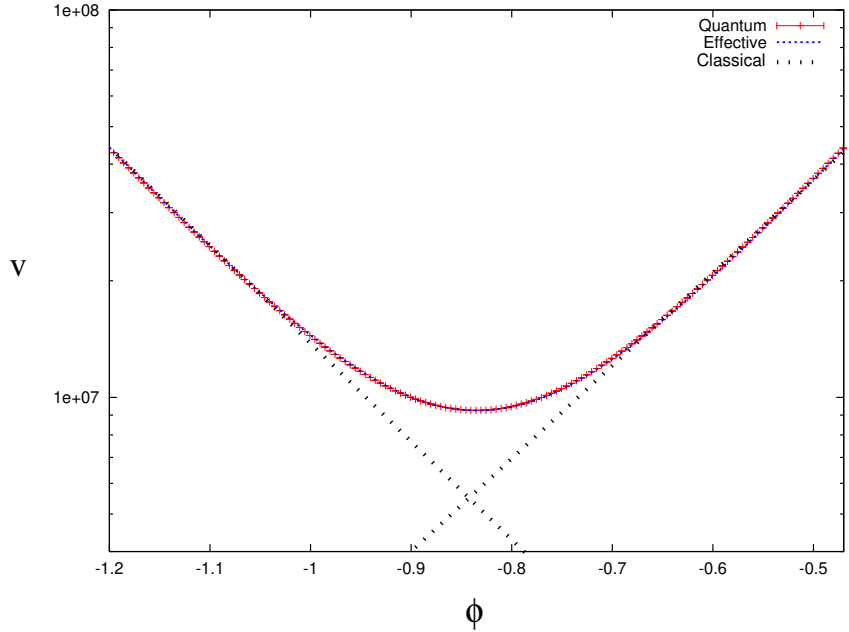}
}
\caption{In this figure, we show the trajectory of the expectation value of the volume operator from the numerical simulations of the quantum difference operator (\ref{quantum_difference_equation}) which illustrates a non-singular evolution of the universe across a quantum bounce. The trajectories from the effective dynamics of LQC and the classical GR are depicted as well in order to make a comparison with the quantum evolution. }
\label{effective-vs-numerical}
\end{figure}

The  quantum Hamiltonian constraint equation (\ref{quantum_difference_equation}) prescribes the quantum evolution of the physical states with respect to the emergent time $\phi$. Unlike its counterpart in the Wheeler-DeWitt theory which is a differential equation on a continuum background spacetime \cite{Ashtekar:2006wn,Ashtekar:2007em}, Eq. (\ref{quantum_difference_equation})  is a second order difference equation with a uniform spacing in the volume.  It naturally divides the space of the physical states into a set of superselection sectors with support on the lattices: ${\cal L}_{\pm \epsilon} = \{ \nu = \pm (4n + \epsilon)\}$ with $n \in \mathbb{N}$ and $\epsilon \in (0,4]$ and the physical states defined in different sectors will not be mixed over time by the evolution equation. Moreover,  using the group averaging techniques \cite{Marolf:1995cn,Marolf:1994wh,Ashtekar:1995zh,Ashtekar:2005dm}, we can define the inner product of the physical states and the relevant Dirac observables, such as  the volume operator at a particular time $\hat v_{\phi_0}$  and the momentum operator  $\hat p_\phi$ \cite{Ashtekar:2006wn}.  These Dirac observables also preserve the superselection sectors and their expectation values under the physical states can be used to extract the physical predictions of the quantum theory. Furthermore,  physical semiclassical states can be constructed precisely in LQC and the numerical simulations of their dynamical evolution under quantum difference equation (\ref{quantum_difference_equation}) lead to quantitative results on the singularity resolution. The evolution of different types of the physical states, starting from the sharply peaked Gaussian states to the highly squeezed and non-Gaussian states,  has been extensively  studied in order to test the robustness of the singularity resolution  \cite{Ashtekar:2006wn,Diener:2014mia,Diener:2013uka,Diener:2014hba}. For the sharply peaked Gaussian states, it has been found that these states remain peaked in the backward evolution during which the volume of the universe shrinks. At very small spacetime curvature, the evolution of the expectation value of the volume operator coincides with  the classical trajectories to a great accuracy. Deviations of the quantum evolution from the classical trajectories occur at high spacetime curvature where the expectation value of the volume operator in the quantum theory bounces at the critical energy density $\rho_c\approx0.41\rho_\mathrm{Pl}$. After the bounce, the volume of the universe starts to increase again. The quantum geometry effects become dominant only in a small neighborhood of the bounce which connects  the  contracting branch with the expanding one.  This feature of the quantum dynamics in LQC can also be precisely captured by the effective dynamics of LQC based on a continuum spacetime description as will be discussed in  detail in Sec. \ref{sec:effective_dynamics}.  In Fig. \ref{effective-vs-numerical},   with the massless scalar field as an emergent time,  we compare the trajectory of the expectation value of the volume operator with those from the effective dynamics of LQC as well as the classical GR. It is obvious from the figure that the effective dynamics gives an accurate approximation of the expectation value of the volume operator from the quantum dynamics throughout the whole evolution of the universe, including in particular the Planckian regime. In addition to the highly peaked states,  the quantum bounce is also found to be a robust feature for the widely spread or squeezed states for which the bounce will occur at a lower maximum energy density than $\rho_c$ \cite{Diener:2014mia,Diener:2013uka}, as well as different  quantization prescriptions in LQC \cite{Martin-Benito:2009htq,MenaMarugan:2011me} and even when more features from full LQG are incorporated into the construction of the model (see Sec. \ref{sec:alternative_cosmological_model} for more details).

The robustness of the singularity resolution has also been further confirmed using analytical solutions in the solvable loop quantum cosmology (sLQC) \cite{Ashtekar:2007em}. This model is obtained by recasting LQC in the spatially-flat FLRW universe with a massless scalar field in the $b$ representation. Due to the availability of  the analytical expressions of the physical states,  one can explicitly compute the expectation values of the volume and the energy density operators for an arbitrary physical state.  It turns out that the expectation values of the volume operator have a non-zero minimum which implies the existence of the bounce. Furthermore, the computation of the expectation values of the energy density reveals an upper bound exactly equal to  $0.41\rho_\mathrm{Pl}$ which is exactly the same as the maximum energy density encountered in the numerical simulations and the effective dynamics. In addition to confirming the robustness of the bounce for an arbitrary state in the physical Hilbert space of LQC, sLQC also plays important roles in various aspects in the investigations of the related subjects, such as understanding the evolution of quantum fluctuations across the bounce \cite{Corichi:2007am,Kaminski:2010yz,Corichi:2011sd,Corichi:2011rt}, computing the quantum probabilities for the occurrence of the bounce in LQC \cite{Craig:2013mga} and singularities in Wheeler-DeWitt theory \cite{Craig:2010vf} as well as exploring  the relations between LQC and the Wheeler-DeWitt theory \cite{Ashtekar:2007em}. Here let us note that the consistent quantum probability for singularity to occur in the above model in LQC turns out to be zero while the bounce has the probability of unity \cite{Craig:2013mga}. On the other hand, the probability for singularity to occur in Wheeler-DeWitt quantization of the same model is unity \cite{Craig:2010vf}.

Before moving onto the next subsection, a few remarks are in order.  Firstly, the mathematical structure of LQC is similar to the polymer quantization of a non-relativistic quantum mechanical system which is implemented  by  realizing the Weyl algebra in the kinematic Hilbert space. The latter is unitarily  inequivalent to  the Schr\"odinger representation in the sense that one of the canonical pair is essentially discrete in the polymer representation and there is no operator corresponding to its conjugate variable. In this way, the von-Neumann theorem is bypassed, resulting in the essential difference between the polymer quantization and the Schr\"odinger quantization  \cite{Ashtekar:2002sn}. Secondly,  LQC adopts a hybrid quantization approach, that is,  the gravitational part of the classical  Hamiltonian constraint is loop quantized while the matter sector is quantized in the Schr\"odinger representation. Therefore,  these two sectors are essentially not treated on the same footing. Since deep in the Planck regime, the matter field also lives on the quantum discretized spacetime, the study of the quantum geometry effects on the matter sector is required for a consistent description of the quantum dynamics. Previous works on the polymer quantization of the scalar field and its cosmological implications can be found  in a series of papers \cite{Hossain:2009ru,Kreienbuehl:2013toa,Hassan:2014sja}.  Further, to obtain the final expression of the quantum difference equation, a unique factor ordering has been chosen for the quantum Hamiltonian constraint operator. There are different choices of the factor ordering as discussed in \cite{Ashtekar:2006uz}. The resolution of the curvature singularity is found to be robust against such a choice. Finally, while in this section we have considered the
$\bar \mu$ scheme for loop quantization arising from the choice \eqref{mu_bar_scheme}, in early literature of LQC a different choice was made where $\mu$ is a constant. It turns out that such a choice runs into various problems, such as with the infra-red limit of the theory \cite{Corichi:2008zb} and occurrence of bounce at a scale which depends on the fiducial cell \cite{Ashtekar:2006wn}. In fact, if one considers any other variation of \eqref{mu_bar_scheme} where $\bar \mu$ is a function of triad only, such as in lattice refined models considered in \cite{Bojowald:2006qu}, one finds that the quantization scheme is severely limited phenomenologically when one considers different matter satisfying null energy condition \cite{Corichi:2008zb}. In this sense, the $\bar \mu$ scheme turns out to be a unique favored choice. This conclusion is also supported by studies on the  stability of the quantum difference equation \cite{Brizuela:2011dv,Singh:2012zc} and the unique factor ordering in the continuum limit of LQC \cite{Nelson:2008vz}.

\section{The effective dynamics of isotropic model}
\lb{sec:effective_dynamics}
\renewcommand{\theequation}{5.\arabic{equation}}\setcounter{equation}{0}

In the previous section we saw that the quantum evolution in LQC is governed by a non-singular quantum difference equation which couples the wavefunction in uniform steps of four Planck volumes in the isotropic model. In general, extracting physical predictions using this equation requires supercomputing resources. However, it is possible to obtain an effective spacetime description\footnote{Unlike the usage of this term ``effective'' in standard quantum field theory where one integrates out high energy modes, in the effective spacetime description of LQC one retains quantum geometric Planck scale effects. The resulting differential equations from the effective Hamiltonian in the $\bar \mu$ scheme of LQC are {\it{not}} classical unless one probes small spacetime curvature regime.}  capturing this dynamics for a suitable choice of coherent states using geometric formulation of quantum mechanics. At first sight it may seem naturally puzzling on how can one capture the non-perturbative quantum gravity effects resulting from quantum geometry in a continuum effective spacetime description. The central result from which one can derive this description is that Schr\"odinger equation is
nothing but the Hamilton's equation in the quantum phase space \cite{kibble,heslot}. While a detailed discussion of this approach is beyond the scope of this Chapter, we provide a brief glimpse of the underlying idea and refer the reader to \cite{ kibble, heslot, strochhi,schilling1, schilling2, brody} for more details.

Using geometrical formulation of quantum mechanics, one can view the Hilbert space as a quantum phase space with a symplectic form $\Omega_Q$ defined via the imaginary part of the Hermitian inner product. The symplectic form allows us to define Poisson brackets and Hamiltonian vector fields. Given a self-adjoint operator $\hat O$, its Schrodinger vector field can be defined as $X_{\hat O} (\psi) := -(i/\hbar) \hat O \psi$. If $\hat O$ is a Hamiltonian operator, this is a Schrodinger equation.  It turns out that the expectation value $\langle \hat O\rangle$ also generates Hamiltonian vector field which is exactly the same as the one generated by $X_{\hat O}$. This implies that if one is interested in obtaining the dynamical evolution of a quantum system determined by a self-adjoint Hamiltonian then that can be equivalently obtained from the Hamilton's equations using expectation values of the Hamiltonian operator. If one considers two arbitrary self-adjoint operators $\hat A$ and $\hat B$ with expectation values $\bar A := \langle \psi| \hat A | \psi \rangle$ and $\bar B := \langle \psi| \hat B | \psi \rangle$ then it is straightforward to show that the Poisson bracket of $\bar A$ and $\bar B$  equals
$\{\bar A, \bar B\}_{\Omega} =  \langle \frac{1}{i \hbar} [\hat A, \hat B] \rangle$. Using which one can show that $\frac{\mathrm{d}}{\mathrm{d} t} \langle \hat A \rangle = \{\bar A, \bar H\}_{\Omega_Q}$.
Time evolution of the expectation values of operator $\hat A$ can be deduced from the Poisson bracket on the quantum phase space. It is to be noted that $\bar H$ is not a classical entity but is the expectation value of the quantum Hamiltonian operator. It is interesting that one can view the Schr\"odinger evolution as resulting from a Hamilton's equation in the quantum phase space.

In LQC the effective spacetime description can be obtained by implementing above techniques and then obtaining an effective Hamiltonian. There are two ways this can be achieved. The first method is based on introducing a coordinate system on the infinite dimensional quantum phase space using
expectation values of quantum operators for phase space variables and their products and higher order moments. Since dynamics is encoded in an infinite number of coupled non-linear differential equations one uses a truncation to a certain order in moments to obtain an approximate evolution of the expectation values \cite{Bojowald:2005cw}. The second method \cite{Willis_thesis,Taveras:2008ke}, which has been widely tested using numerical simulations \cite{Diener:2014mia,Diener:2013uka,Diener:2014hba}, is based on finding a reliable embedding of the classical phase space into the quantum phase space. Finding this embedding which is preserved by the Hamiltonian flow is a non-trivial task which requires a judicious choice of semi-classical states. In LQC this embedding has so far been found for the spatially-flat isotropic universe sourced with a massless scalar field. In
the following we consider the effective Hamiltonian for the massless scalar case and generalize the setting to arbitrary matter assuming the validity of the effective Hamiltonian approach.

\subsection{The modified Friedmann and Raychaudhuri equations in the $k=0$ FLRW universe}

In a spatially-flat FLRW universe, the effective Hamiltonian constraint has been derived in the case of the massless scalar field  using the using the embedding method. In terms of
$v$ and $b$ variables introduced in Sec. \ref{sec:spatially_flat_FLRW}, it turns out to be \cite{Taveras:2008ke}
\bq
\mathcal H=-\frac{3v\sin^2\left(\lambda b\right)}{8\pi G\gamma^2\lambda^2}+\frac{p^2_\phi}{2v}
\eq
where the matter content is taken to be a massless scalar field. Note that the same Hamiltonian can also be obtained from the classical Hamiltonian constraint (\ref{classical_FLRW_Hamiltonian_constraint}) by using the thumb rule $b^2\rightarrow\sin^2\left(\lambda b\right)/\lambda^2$ which as a cautionary remark only holds for the spatially-flat FLRW universe.  The modified Friedmann and Raychaudhuri equations can be derived from the Hamilton's equations
\bq
\dot v=\frac{3v}{2\lambda \gamma}\sin\left(2\lambda b\right),\quad \quad \dot b=-\frac{3\sin^2\left(\lambda b\right)}{2\gamma\lambda^2}-\frac{2\pi G \gamma p^2_\phi}{v^2}.
\eq
Using the Hamiltonian constraint and the equation of motion for the volume, it is straightforward to obtain the modified Friedmann equation \cite{Ashtekar:2006wn,Singh:2006sg}
\bq\label{eq:modfried}
H^2=\frac{8\pi G}{3}\rho\left(1-\frac{\rho}{\rho_c}\right) .
\eq
Here $\rho$ as before denotes the energy density of the matter content, which in the present case is $\rho = p^2_\phi/(2 v^2)$,  $\rho_c$ is the maximum density equal to $\rho_c =3/8\pi G \lambda^2\gamma^2\approx0.41$ for $\gamma=0.2375$ which is determined by the black hole thermodynamics in LQG \cite{Meissner:2004ju}. If we assume the validity of the effective Hamiltonian constraint for all  the matter content, the quantum bounce takes place at  the maximum energy density $\rho_c$ where the Hubble rate vanishes. Before the bounce, there appears a contracting phase which also has the classical limit in the distant past when the energy density becomes much less than the Planck density.  The evolution of the universe filled with a massless scalar field  is symmetric with respect to the quantum bounce  and the Hubble rate is bounded throughout the  evolution. Moreover, the magnitude of the Hubble rate attains its maximum $H^2_\mathrm{max}=1/4\gamma^2\lambda^2$ at $\rho=\rho_c/2$. In the regime $\rho_c/2<\rho\le \rho_c$, the universe enters into a super-inflationary phase with $\dot H>0$.

The Raychaudhuri equation can be similarly obtained from the equation of motion for $b$, which turns out to be
\bq\label{eq:mod-ray}
\frac{\ddot a }{a}=-\frac{4\pi G}{3}\rho\left(1-4\frac{\rho}{\rho_c}\right)-4\pi G P\left(1-2\frac{\rho}{\rho_c}\right),
\eq
where $P$ denotes the pressure of the scalar field. It is obvious that when the energy density is far less than the Planck density, both the modified Friedmann and Raychaudhuri equations asymptote to their classical counterparts. Finally, combining the modified Friedmann and Raychaudhuri equations, it is straightforward to obtain the energy conservation law
\bq
\dot \rho+3H\left(\rho+P\right)=0,
\eq
which takes the same form as its classical counterpart. The validity of these effective equations has been tested rigorously using high performance computing methods \cite{Diener:2014mia,Diener:2013uka,Diener:2014hba}, including extension to anisotropic models \cite{Singh:2018rwa,Diener:2017lde}. It has been found that if one starts with states sharply peaked on the classical trajectory in the expanding branch and evolves them backwards then states retain their peakedness properties throughout the evolution and one recovers a classical pre-bounce universe before the bounce. If instead one considers a very quantum state then the state retains its quantum character during the evolution through the bounce. Further, the energy density at bounce in such a case decreases \cite{Diener:2014hba} by exactly the same amount as predicted by exactly solvable model of spatially-flat universe in LQC sourced with a massless scalar field \cite{Corichi:2011rt}. For this particular model, bounce at lower density due to quantum fluctuations can also be captured by a modified Friedmann equation resembling \eqref{eq:modfried} \cite{Ashtekar:2015iza}.

We conclude this subsection with a few remarks. First let us note that the modifications to the Friedmann dynamics are non-perturbative in nature and can not be captured by introducing a finite number of higher curvature terms. Indeed these arising from non-local effects require a non-local action. Such an effective action has been obtained  in the Palatini framework which very accurately results in the modified Friedmann equation \ref{eq:modfried} \cite{Olmo:2008nf}. Similarly an effective action has been obtained for the variants of LQC which result in much more complex modified Friedmann dynamics using Palatini framework \cite{adria}. Second, in the above treatment we have excluded inverse volume modifications in the effective dynamics. While they are not applicable for non-compact models, they still play a negligible role compared to holonomy modifications for compact spatially-flat models if  bounce happens at volumes much greater than Planck volume \cite{Ashtekar:2006wn}. However, inverse volume modifications can play a role in  singularity resolution in the presence of spatial curvature. In fact, they can lead to singularity resolution just by themselves \cite{Singh:2003au}, allow probing the nature of non-singular quantum spacetime near and at the zero scale factor \cite{meysam}, and are also important for bounds on anisotropic shear in Bianchi-IX model \cite{Gupt:2011jh}. However there is a caveat to be noted in the effective description. In the regime when inverse scale factor become important, fluctuations of states are expected to be large. But, as mentioned above, for states with large fluctuations modified Friedmann equation surprisingly turns out to be valid but with a lower bounce density. While this result has so far been found for spatially-flat model, note that for small scale factors spatial curvature term is much smaller than the matter energy density. While this does not provide any direct evidence of the validity of modified Friedmann like equations in the deep Planck regime for more general cases than as discussed in Ref. \cite{Ashtekar:2015iza}, it does indicate the domain of validity of
modified Friedmann dynamics may be larger than expected. Nevertheless, in the following it is important to recall the caveat of assuming the validity of effective dynamics in the entire regime.

\subsection{Generic resolution of singularities}

We have seen so far that quantum geometry effects in LQC can successfully resolve the big bang singularity for isotropic universe sourced with a  massless scalar field. If we assume that the effective Hamiltonian is valid for different types of matter in all regimes it is easily seen that the modified Friedmann dynamics results in a bounce in the Planck regime if the matter satisfies the weak energy condition. However, as discussed in Sec. 3, cosmological singularities can be more general than the big bang/crunch singularities. This results in a pertinent question: whether loop quantum effects resolve all of the spacelike singularities? To answer this question it is important to understand whether loop quantum geometric effects always bound the spacetime curvature, or if there are exceptions. What happens to the fate of strong and weak singularities? Finally, one would like to answer one of the most important questions on singularity resolution: Are loop quantum spacetimes geodesically complete?

While the geodesic extendibility tells us whether the event at which curvature diverges is a physical singularity or not, the strength of the singularities as discussed in Sec. 3.1 tells us
whether an in-falling observer or detector is completely annihilated by the tidal forces. These complimentary approaches together help us understand a more complete picture of physics of singularities. In classical cosmological context, geodesic incompleteness signals strong singularities and vice versa. Similarly, in the classical theory a passable curvature divergent event in geodesic evolution in a cosmological model is linked to a weak singularity. Note that quantum geometric effects can in principle resolve the strong singularity but for a non-singular description it suffices if they just convert it to a weak one.\footnote{It is possible that such a passage from a curvature divergent event in the effective description may result in additional complexities for the effective description. However, as everywhere else in this Chapter we assume the validity of effective description in all regimes.}

The fate of geodesic completeness, and of strong and weak singularities has been explored in detail in LQC using the effective spacetime description \cite{Singh:2014fsy}. Let us recall that using an exactly solvable model of LQC one can show that there is a non-zero lower bound on the expectation value of the volume operator and a universal upper bound on  expectation value of the energy density operator for the states in the physical Hilbert space \cite{Ashtekar:2007em}.   At the level of effective spacetime description, these results generalize to all types of matter but do not exclude the possibility of divergence in curvature components due to divergence in pressure at a finite density and scale factor. This is in addition to the boundedness of the Hubble rate which is directly responsible for lack of breakdown of geodesic  evolution and resolution of strong singularities. The quantum geometric effects which manifest themselves in boundedness of relevant observables in the quantum theory, such as volume and energy density, results in an effective spacetime which is geodesically complete and free of strong singularities.
It has been proved that for arbitrary matter content there can be no strong curvature singularities in the isotropic models \cite{Singh:2009mz} as well as in presence of spatial curvature \cite{Singh:2010qa} and anisotropies which includes Bianchi-I \cite{Singh:2011gp}, Bianchi-II \cite{Saini:2017ipg} and Bianchi-IX models \cite{Saini:2017ggt}. These results also extend to the Kantowski-Sachs spacetime which in absence of matter captures the interior of the Schwarzschild black hole \cite{Saini:2016vgo}. In all these spacetimes, LQC results in geodesic completeness. In addition these results have been checked to be robust in modified versions of LQC for the spatially-flat isotropic model \cite{Saini:2018tto}.
An interesting result from these investigations is that LQC permits weak curvature singularities.  At these singularities spacetime curvature in LQC can indeed diverge. This can be easily seen via the modified Raychaudhuri equation \eqref{eq:mod-ray}. While the quantum geometry effects in LQC universally bound the energy density, they do not bind the pressure. Therefore, for an appropriate choice of equation of state where the  pressure can diverge at finite energy density, the spacetime curvature can diverge in LQC. Recalling the types of singularities from Sec. 3.1, one finds that type-II singularities are not resolved in LQC \cite{Singh:2009mz}. Similarly, type-IV singularities which occur because of divergence in derivatives of spacetime curvature are also not resolved. Note that both of these singularities are weak singularities and thus harmless for geodesic evolution. In addition to these results, various models have been explored assuming phenomenological equation of state which permits study of various types of cosmological singularities \cite{Sami:2006wj,Samart:2007xz,Naskar:2007dn,Cailleteau:2008wu,Wu:2008db}, in which the fate of various types of singularities has been studied. These investigations confirm that all types of strong curvature singularities are resolved in LQC.  Indeed it can happen that strong curvature singularities are converted to weak ones because of quantum geometry effects making them harmless \cite{Singh:2009mz}. An open question in this arena is whether singularity resolution which is robustly seen for isotropic and anisotropic models also applies in the presence of inhomogeneities.

\subsection{The inflationary scenario in LQC}

The inflationary paradigm is one of the cornerstones in the  modern cosmology. It does not only resolve some long-standing puzzles in the standard big-bang model  but also provides a mechanism for the development of large scale structure in the universe. However, the inflationary paradigm itself is not past-complete as pointed out in the incompleteness theorem by Borde, Guth and Vilenkin \cite{Borde:2001nh} and  the inflationary spacetime in a spatially-flat FLRW universe would inevitably encounter a singularity in the backward evolution as long as the null energy condition is satisfied \cite{Hawking-Ellis}.  To resolve the big-bang singularity in the inflationary paradigm, a straightforward phenomenological model in LQC is to make use of the effective dynamics by adding an inflationary potential into the effective Hamiltonian constraint. Although  with the inclusion of the inflationary potential,  the scalar field can no longer serve as a matter clock due to its non-monotonicity, the additional matter clocks (also called reference fields), such as the dust fields or the massless Klein-Gordon fields can be  taken into account, resulting in a ``two-fluid" system \cite{Giesel:2020raf}. The loop quantization of such a system is implemented in the reduced phase space where the Dirac observables with respect to the matter clock are explicitly constructed.  The quantization proceeds in a similar way with the $\bar \mu$ scheme as described in Sec. \ref{sec:isotropic_model}, leading to the quantum difference equations with the same non-singular structure as in LQC.  It turns out that the effects of the reference fields on the background evolution of the universe  can be tuned as small as possible when the magnitudes of their energy densities are chosen to be sufficiently small \cite{Giesel:2020raf}. Therefore, in the following, we simply add the inflationary potential directly to the effective Hamiltonian constraint as a phenomenological model to consider the effects of the inflationary potential in an LQC universe without taking into account any additional effects from  the reference fields. This is also the common practice in the literature to study the extension of the inflationary paradigm to the Planck regime in the framework of LQC.  Moreover, numerous inflationary models have been discussed extensively in LQC, such as single field inflation \cite{Singh:2006im,Ashtekar:2009mm,Corichi:2010zp,Mielczarek:2010bh,Ashtekar:2011rm,Linsefors:2013cd,Bonga:2015kaa,Ashtekar:2015dja,Ashtekar:2016wpi,Zhu:2017jew,Shahalam:2018rby,Sharma:2018vnv},  multi-field inflation \cite{Ranken:2012hp}, tachyonic inflation \cite{Xiong:2007ak}, inflation with non-minimally coupled scalar field \cite{Artymowski:2012is}, in presence of anisotropies \cite{Gupt:2013swa}, spatial curvature \cite{Gordon:2020gel,Dupuy:2016upu, Dupuy:2019ibu}, and warm inflationary scenarios which include radiation dissipation \cite{Motaharfar:2021gwi}.

Since Planck 2018 data prefers single field inflationary model with a plateau in a spatially-flat universe \cite{Planck:2018jri}, we will focus on this simplest model of inflation in the rest of this subsection. The effective Hamiltonian for a massive scalar field minimally coupled to gravity in LQC takes the form
\bq
\mathcal H=-\frac{3v\sin^2\left(\lambda b\right)}{8\pi G\gamma^2\lambda^2}+\frac{p^2_\phi}{2v}+vU,
\eq
where $U$ stands for the inflationary potential. Using the effective dynamics generated by the above background Hamiltonian constraint, various inflationary potentials with their phenomenological implications on the background evolution of the LQC universe have been investigated in the last decade, such as the chaotic potential,  the fractional monodromy potential, the Starobinsky potential, the non-minimal Higgs potential and the $\alpha$ attractor. Among them, the most studied examples are the chaotic potential and the Starobinsky potential. It has been found that although the bounce can be dominated either by the kinetic energy or the potential energy of the inflaton field, only the kinetic-energy-dominated bounce turns out to be relevant to the observations which requires around $60$ inflationary e-foldings. The potential-energy-dominated bounce can exhibit distinct properties of the inflationary phase with a different choice of the potential. For example, with the chaotic potential, the  potential-energy-dominated bounce can lead to excessively long period of the inflationary phase while with the Starobinsky potential, there would be even no sustained period of the inflationary phase if the bounce  is dominated by the potential energy. As a result, various investigations have been directed to study the properties of the universe with a kinetic-energy-dominated bounce by using numerical simulations as well as the analytical approximations. It has been found that such a universe undergoes several characteristic stages evolving from the quantum bounce in the Planck regime \cite{Zhu:2017jew}. These stages are featured by the distinct behavior of the equation of state of the inflaton field. The bounce, including the super-inflationary phase, is located in the Planck regime where the equation of state is very close to unity due to the negligible potential energy. It is then followed by the transition phase in which the equation of state  quickly drops from positive to negative unity in a  couple of e-foldings. Afterwards,  the slow-roll phase starts and its duration is directly determined by the initial value of the inflaton field at the quantum bounce. In addition to the numerical results, the analytic expressions of the background solutions have also been studied for each phase particularly for the chaotic and the Starobinsky potentials which serve as the basis for an analytical investigation of the primordial power spectra in LQC later on \cite{Zhu:2017jew}.

\begin{figure}
{
\begin{center}
\includegraphics[width=12cm]{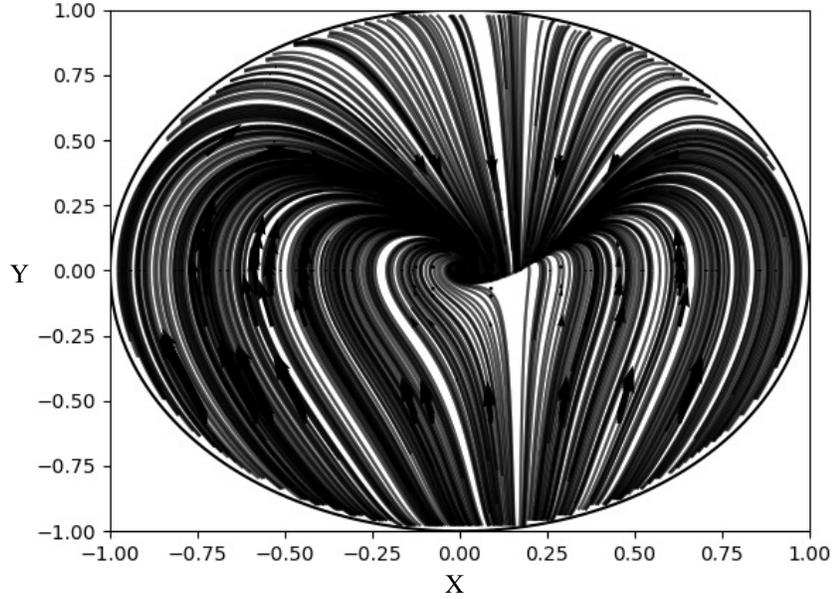}
\end{center}
}
\caption{This figure shows a 2-dimensional phase space portrait  of the qualitative dynamics of the universe filled with a massive scalar field with mass $m$ for the Starobinsky potential in LQC. The x-axis is $X = \chi_0 (1 - \exp(-\sqrt{16 \pi G/3} \phi))$ with $\chi_0 = \sqrt{\frac{3 m^2}{32 \pi G \rho_c}}$. The y-axis is $Y = \dot \phi/\sqrt{2 \rho_c}$. It  illustrates the generic evolution of the physical solutions from the quantum bounce to the reheating phase. The arrows in the figure denote the time flow in the forward evolution. All the trajectories starting from the unit circle which represents the quantum bounce first tend to the inflationary separatrices and then circling around the origin, signifying the decaying of the energy density in  the reheating phase.  }
\label{phase_portrait}
\end{figure}

On the other hand, to fully explore the parameter space of the initial conditions at the quantum bounce which is essentially one-parameter space determined by the value of the inflaton field, the qualitative dynamics of the inflationary model in LQC is analyzed by casting the universe filled with an inflaton field into an autonomous system described by a closed set of first-order differential equations \cite{Singh:2006im}. The two-dimensional phase space portraits can be obtained from this set of equations with x/y axis proportional to the square root of the potential/kinetic energy of the inflaton field. A representative example of the phase space portraits is illustrated in Fig. \ref{phase_portrait} in which we use the Starobinsky potential as an example. All the trajectories are confined within the unit circle which represents  the quantum bounce  at the maximum energy density  in LQC.
From the above phase space portraits, one can intuitively see that ``most" of the initial conditions can lead to the slow-roll phase before reaching the reheating phase located at the center of the plot. In LQC, one can unambiguously compute the probability for the occurrence of the desirable slow-roll phase which is defined as the one with enough e-foldings to allow the  pivot mode to exit the Hubble horizon during the slow-roll phase. The key point is the additional structure provided by the quantum bounce in LQC. As is well-known, the problem with having a definite probability for the occurrence of the desirable slow-roll phase in classical GR is tied up to the fact that the Liouville measure on the gauge fixed surfaces $H=H_0=const$ fails to naturally descend to a measure in the space of physically distinct solutions. As a result, different choices of the gauge-fixed surface, namely different values of $H_0$, can even lead to opposite results on the probability \cite{Schiffrin:2012zf}. In contrast, there exists a canonical choice of the time instant, i.e. the bounce surface, which is endowed with a distinguished physical measure yielding a finite physical volume of the available parameter space at the bounce. It can be shown that for the chaotic potential, the possibility for the desirable slow-roll phase to not  happen in LQC is less than three in a million \cite{ Corichi:2010zp, Ashtekar:2011rm,Corichi:2013kua}. Similar results have been obtained for other versions of LQC obtained from treating Euclidean and Lorentzian parts of the Hamiltonian constraint independently resulting in mLQC-I and mLQC-II \cite{Li:2018fco, Li:2019ipm}. No such are available for the Starobinsky potential. This is because  the physical volume of the parameter space turns out to be  infinite,  and hence no estimates on the possibility for the desirable slow-roll to happen can be found.

\subsection{The matter bounce and the ekpyrotic scenarios in LQC}

In this subsection, we briefly summarize the results from the ekpyrotic and the matter-ekpyrotic bounce scenario in LQC. The matter bounce scenario and the ekpyrotic scenario are actually two separate ansatz to address different issues encountered in the construction of alternative phenomenological models to the inflationary paradigm.\footnote{Note that as in the case of inflationary potentials  studied in LQC, the ekpyrotic potential is not derived from LQG and is used in the following discussion only as a phenomenological model to understand the role of quantum geometric effects in presence of ekpyrosis which here means a period where equation of state is greater than unity.}
 The matter bounce scenario \cite{Brandenberger:2012zb} aims to produce a scale-invariant power spectrum of the linear perturbations during the contracting phase of the universe in a period of vanishing equation of state of the matter field which is dual to the inflationary phase in the expanding branch \cite{Wands:1998yp}. In GR, the most difficult part for such a scenario is the realization of a non-singular bounce, which is usually achieved by introducing an exotic matter content which violates null energy condition  or modified gravitational sector \cite{Cai:2008ed,Brandenberger:2009yt,Lin:2010pf,Battarra:2014tga,Sahoo:2019qbu}. In contrast,
the generic resolution of the big bang singularity due to the quantum geometry effects in LQC provides a natural framework to investigate the physical  implications of such a scenario without altering either the geometric or the matter sectors. There is no need to introduce any exotic matter or fine tuning of initial conditions and one can simply consider a massless scalar field or a dust field in the bouncing universe in LQC.
However, it turns out that the magnitude of the power spectra produced in such a simple setting is proportional to the maximum energy density in LQC which is of the Planck scale \cite{Wilson-Ewing:2012lmx}. Hence, considering only a matter field with vanishing equation of state in LQC is not sufficient to produce a scale-invariant power spectrum with the observationally consistent magnitude.

On the other hand, the ekpyrotic scenario provides an alternative explanation on the origin of the universe \cite{Khoury:2001wf,Khoury:2001bz}. In this framework, our universe is supposed to exist on a brane interacting with other branes in a higher dimensional bulk. The big bang/big crunch singularities correspond to the collision of branes and the  the universe is supposed to undergo cycles of contracting and expanding phases. This evolution is captured by a moduli field which measures the inter-brane distance. The inter-brane potential has a peculiar shape with a sharp negative well. When the moduli field is in this well its equation of state can become much greater than unity. In this case ekpyrotic matter energy density dominates the dynamics even in presence of anisotropies which scales as $\sigma^2 \propto a^{-6}$.
In principle this phenomena can alleviate the chaotic approach to singularities and the Belinski-Khalatnikov-Lifshitz (BKL) instability \cite{Belinsky:1970ew}. But the problem with conventional ekpyrotic scenarios is the presence of
big bang/big crunch singularity. In LQC,  the ekpyrotic scenario was first discussed in \cite{Singh:2006im,Bojowald:2004kt} where singularity resolution was achieved. It was noted that even with singularity resolution there exist inherent problems, which are independent of LQC,  with lack of turn-around of the moduli field along with that of the scale factor at the bounce. Later it was found that in order to form a cyclic universe, in addition to the ekpyrotic field, another matter field or anisotropies must be also taken into account \cite{Cailleteau:2009fv}. Extensive numerical simulations have been performed in the Bianchi-I model in LQC with ekpyrotic and an ekpyrotic-like potential which confirm resolution of singularity in these models due to non-perturbative effects \cite{McNamara:2022dmf}.

Since two scenarios discussed above address different problems encountered in the construction of a feasible alternative paradigm to the inflationary one, it is useful to study them together which results in the matter-ekpyrotic bounce scenario \cite{Cai:2012va,deHaro:2015wda}. In LQC, the physical implications of the matter-ekpyrotic bounce scenario on the background dynamics as well as the primordial power spectra have been investigated in \cite{Cai:2014zga,Haro:2017mir,Li:2020pww}.  In these works, gravity is minimally coupled to two fluid, namely a dust field which plays the role of producing the scale-invariant power spectra in the contracting phase and the ekpyrotic field which isotropizes the bounce.
The contracting branch is thus divided into two stages, the dust-dominated phase far from the bounce and the ekpyrotic phase near the bounce which helps remove the reliance of the magnitude of the power spectra on the maximum energy density at the bounce in LQC.
Using the dressed metric approach to perturbations it has been found that the primordial power spectrum of comoving curvature perturbations is almost scale invariant for the modes which exit the horizon in the matter-dominated phase \cite{Li:2020pww}. Analysis of the spectral index and observational constraints show that refinements in this model are necessary. Finally, let us note that while ekpyrotic models can be successfully embedded in LQC the same can not be said if we generalize to another variant of LQC  \cite{Li:2021fmu} (see Sec. 7.1 for a discussion).

\section{Loop quantization in presence of anisotropies and inhomogeneities}
\lb{quantization-anisotropies_inhomogenities}
\renewcommand{\theequation}{6.\arabic{equation}}\setcounter{equation}{0}

In this section, we overview the loop quantization of the Bianchi-I model in the $\bar \mu$ scheme and the polarized Gowdy models with the inhomogeneities as infinite degrees of freedom. When extending the techniques developed for the isotropic model to the cases in presence of  anisotropies such as Bianchi-I spacetime, one has to handle several new difficulties. These include defining a curvature operator in  terms of the holonomies which depend on the directional triads in a non-trivial way, finding a proper set of the dynamical variables that can result in a manageable Hamiltonian constraint operator and etc. More importantly, the resulting Hamiltonian constraint operator should not suffer from the drawbacks encountered in the $\mu_0$ scheme of the isotropic case as discussed earlier. It turns out that there exist two quantization prescriptions, one known as Madrid prescription \cite{Chiou:2007sp, Martin-Benito:2008dfr} and another known as Ashtekar-Wilson Ewing prescription \cite{Ashtekar:2009vc},  which result in the $\bar \mu$ scheme in the isotropic limit. However, the former becomes unviable with $\mathbb{R}^3$ topology \cite{Corichi:2009pp} and also suffers from problematic features at large volumes \cite{bianchi-new}. For the Ashtekar-Wilson prescription, cosmological implications on the non-singular evolution of the Bianchi-I universe can be found in \cite{Singh:2011gp,Gupt:2013swa,McNamara:2022dmf,Gupt:2012vi}.  For the loop quantization of other anisotropic models using same prescription, see   \cite{Gupt:2011jh,Singh:2013ava,Corichi:2016xfy}.

In addition to the spacetimes with anisotropies, it is also equally important to incorporate inhomogeneities as a step towards understanding the quantum geometry effects and investigating the robustness of the singularity resolution in presence of the infinite degrees of freedom. This provides useful insights on the possible properties of the cosmological sector of the full theory.  In the following, we will summarize the main results  of the $\bar \mu$ scheme quantization of the Bianchi-I model and  discuss its phenomenological implications by using the effective dynamics. Then we will move onto a brief summary of the simplest Gowdy $T^3$ which uses the construction of Bianchi-I spacetime.

\subsection{Loop quantization of Bianchi-I spacetime and its effective dynamics}

A detailed construction of the quantum theory of loop quantized Bianchi-I spacetime is addressed in \cite{Ashtekar:2009vc}. Here we outline the basic ideas and procedures of the quantization  which is implemented in a similar way as that of the isotropic model discussed in Sec. \ref{sec:LQC_in_spatially_flat_universe}. The novel features of the quantization come from the presence of anisotropies which result in six degrees of freedom in the classical phase space, namely the directional connection variables $c_i$ and the directional triads $p_i$. Correspondingly, the fundamental variables for the loop quantization are the directional triads and the holonomies of the directional connection variables. Using the same techniques applied in the isotropic case, one can construct the kinematic Hilbert space in the triad representation with the almost periodic functions of the connection as the basis states. Meanwhile, to construct a physically viable Hamiltonian constraint operator, one of the key elements is to find a proper regularization of the field strength operator. In the isotropic case, such an operator is defined on a square $\square_{ij}$ as computed exactly in (\ref{field_strength}). With the physical area of the square shrinking  to the minimal non-zero eigenvalue of the area operator in the homogeneous spacetime, the unique $\bar \mu$ scheme is obtained. In the case of the Bianchi-I spacetime,  the edge length $\bar \mu_i$ in each direction is determined from a correspondence between the kinematic states in LQG and those in LQC which yields the specification of the edge length as
\bq
\lb{mu_bar_bianchi}
\bar \mu_1=\lambda\sqrt{\frac{|p_1|}{|p_2p_3|}}, ~~~\bar \mu_2=\lambda\sqrt{\frac{|p_2|}{|p_3p_1|}},~~~\bar \mu_3=\lambda\sqrt{\frac{|p_3|}{|p_1p_2|}}.
\eq
When anisotropies disappear, the above value of the edge length in each direction reduces to the one used in the $\bar \mu$ scheme in the isotropic case  given in (\ref{mu_bar_scheme}). Another key observation in the quantization procedure is the use of a new set of dynamical variables which can lead to a manageable Hamiltonian constraint operator. It turns out that   instead of using the states $\Psi(p_1,p_2,p_3)$, it is more convenient to work out the action of the Hamiltonian constraint operator on the wave function $\Psi(\lambda_1,\lambda_2,v)$ with
\bq
\lambda_i=\frac{sgn(p_i)\sqrt{|p_i|}}{(4\pi \gamma \lambda l^2_\mathrm{Pl})^{1/3}},\quad  \quad v=2\lambda_1\lambda_2\lambda_3,
\eq
where $i=1,2,3$ and $\lambda_i$ has the same sign as $p_i$. In the absence of the fermions, the physical wave function  $\Psi(\lambda_1,\lambda_2,v)$  should be symmetric under a flip in the orientation of the fiducial triads in each direction, demanding the requirement  $\Psi(\lambda_1,\lambda_2,v)=\Psi(|\lambda_1|,|\lambda_2|,|v|)$, and meanwhile be annihilated by the Hamiltonian constraint operator, leading to a dynamical difference equation whose evolution is again unfolded with respect to the massless scalar field. The exact form of the quantum difference equation  is rather complicated as compared with its isotropic cousin given in (\ref{quantum_difference_equation}). Therefore, we will not explicitly cite it here and  interested readers can refer to Eqs. (3.13)-(3.16) in \cite{Ashtekar:2009vc}.  The quantum difference equation in the $\bar \mu$ scheme for the Bianchi-I spacetime also has the uniform steps in the volume, that is, it  involves only  the physical  states with  the  volumes $v$ and $v\pm4$. For this reason, there  also exist  superselection sectors with support on  lattices ${\cal L}_{\pm \epsilon} $ as in the isotropic case. However, there is no superselection with respect to the first two arguments, namely $\lambda_1$ and $\lambda_2$, of the wave function. Moreover, $\lambda_1$ and $\lambda_2$ which carry the information of the anisotropies do not appear in the coefficients of the quantum difference  equation but only appear in the arguments of the wave function. Under the action of the Hamiltonian constraint operator they get rescaled by factors depending only on the volume.  Owing to this property, it turns out that there exists a natural projection from the physical states in the Bianchi-I Hilbert space to those in the isotropic Hilbert space. When integrating out the anisotropic degrees of freedom, the quantum difference equation of the Bianchi-I model can exactly reduce to the one in the isotropic $k=0$ FLRW model \cite{Ashtekar:2009vc}.  This provides a typical example of how the symmetry reduced quantum dynamics can be obtained from a more general one. Considering the BKL conjecture which asserts that the dynamics of GR near space-like singularities can be well modeled by Bianchi-I cosmology, the projection of the Bianchi-I model to the isotropic FLRW model implies  that the latter may capture some genetic features of the quantum dynamics of the homogeneous and isotropic sector of the full theory. Finally, it can be shown that the quantum Bianchi-I difference equation can be well approximated by the Wheeler-DeWitt differential equation when the quantum geometry effects are negligible. However, in the Planck regime, this approximation fails and two quantum theories lead to  distinct physical predictions as in the isotropic case.

\begin{figure}
{
\begin{center}
\includegraphics[width=12cm]{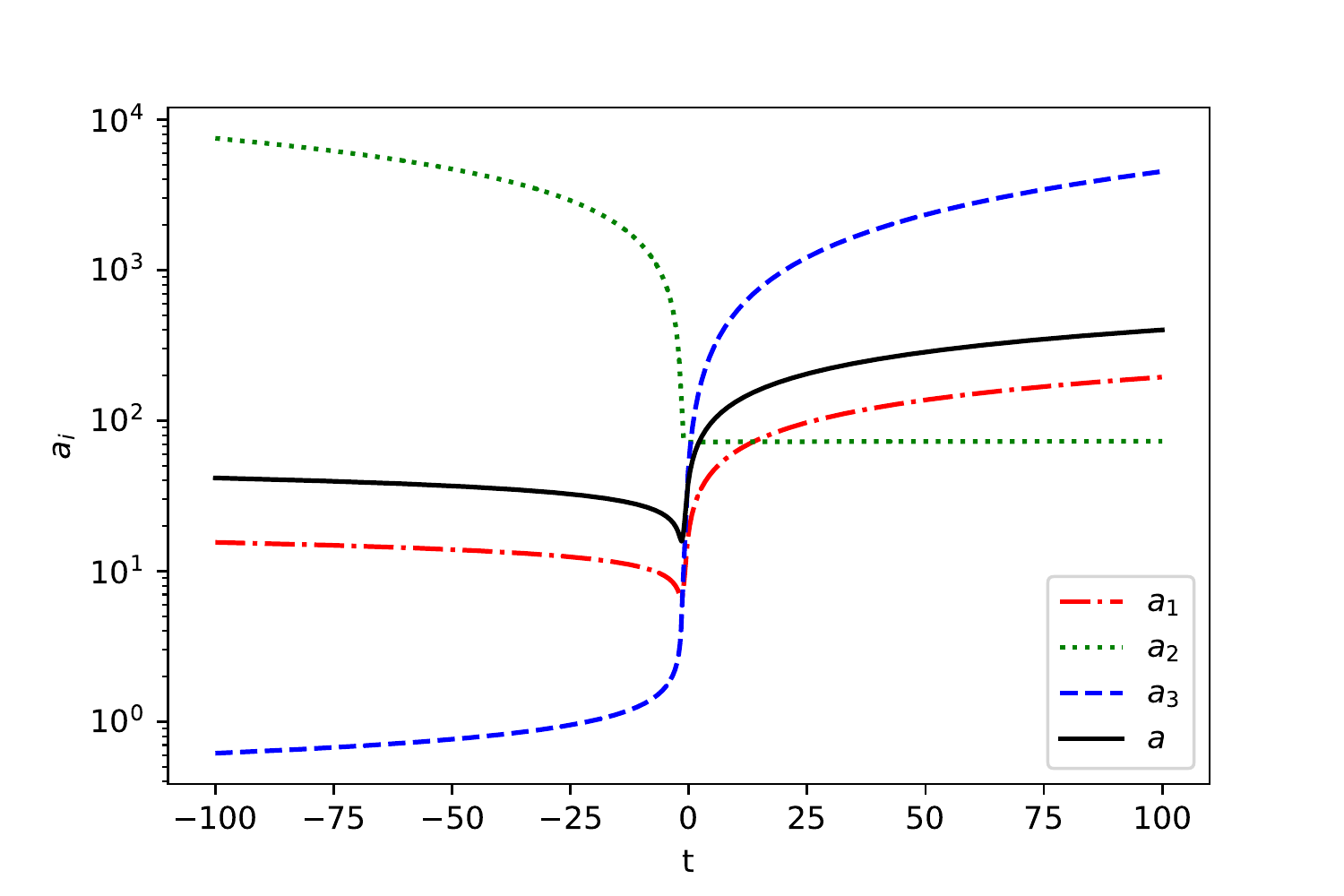}
\end{center}
}
\caption{Non-singular evolution of scale factors in the loop quantized Bianchi-I spacetime is shown for the case of an inflationary potential. The approach to classical singularity is cigar-like with two scale factors decreasing and one increasing. The resulting shape of the bounce is thus not point-like but as a cigar. Singularity is resolved with finite values of energy density and anisotropic shear.}
\label{scalefactors}
\end{figure}

Similar to the isotropic models, the effective description of the quantum dynamics is also shown to be numerically valid for the Bianchi-I model when the background quantum states are highly peaked around the classical trajectories at late times \cite{Singh:2018rwa,Diener:2017lde}. A caveat is that this analysis has been so far performed only for the Madrid prescription in the vacuum case. These semi-classical states remain peaked throughout the whole evolution of the background spacetime and  hence the effective dynamics also faithfully captures the quantum geometric effects in the Planck regime. Therefore, the effective dynamics is  widely used to explore the cosmological implications of the quantum dynamics of  the Bianchi-I universe. The corresponding effective Hamiltonian constraint in the $\bar \mu$ scheme is given explicitly by  \cite{Chiou:2007sp,Ashtekar:2009vc}
\bq
\mathcal H^\mathrm{eff}=-\frac{1}{8\pi G \gamma^2 v}\left(\frac{\sin(\bar \mu_1 c_1)}{\bar \mu_1}\frac{\sin(\bar \mu_2 c_2)}{\bar \mu_1}p_1p_2+\mathrm{cyclic~ terms}\right)+v\rho,
\eq
where $\bar \mu_1$, $\bar \mu_2$ and $\bar \mu_3$ are given in (\ref{mu_bar_bianchi}). In the isotropic limit when $p_1=p_2=p_3$, the above effective Hamiltonian constraint reduces to the one in the isotropic model. Using the Poisson bracket between connection and triad variables, $\lbrace c_i,p_j \rbrace = 8\pi G \gamma \delta_{ij}$ it is straightforward to derive the Hamilton's equations with non-perturbative quantum gravity modifications. For $c_1$ and $p_1$ variables these turn out to be
\begin{align}
\label{triad_flow}
\dot{p_1}=&\frac{p_1}{\gamma\lambda}\left(\sin(\overline{\mu}_2c_2) + \sin(\overline{\mu}_3c_3)\right)\cos(\overline{\mu}_1c_1), \\
\dot{c_1} =& \frac{v}{2\gamma \lambda^2 p_1}\Big[ c_2\overline{\mu}_2\cos(\overline{\mu}_2c_2)\left(\sin(\overline{\mu}_3c_3) + \sin(\overline{\mu}_1c_1)\right) \nonumber \\ & ~~~~~~~~~~ +
c_3\overline{\mu}_3\cos(\overline{\mu}_3c_3)\left(\sin(\overline{\mu}_1c_1) + \sin(\overline{\mu}_2c_2)\right) \nonumber \\
& ~~~~~~~~~~ - c_1\overline{\mu}_1\cos(\overline{\mu}_1c_1)\left(\sin(\overline{\mu}_2c_2) + \sin(\overline{\mu}_3c_3)\right) \nonumber\\
& ~~~~~~~~~~ - \left(\sin(\overline{\mu}_1c_1)\sin(\overline{\mu}_2c_2) + \sin(\overline{\mu}_2c_2)\sin(\overline{\mu}_3c_3) + \sin(\overline{\mu}_3c_3)\sin(\overline{\mu}_1c_1)\right) \Big] \nonumber\\
& ~~~~~~~~~~+ 8\pi G\gamma \frac{\partial v \rho}{\partial p_1} .
\end{align}
Similarly one can obtain equations of motion for other connection and triad variables. In the general case when the anisotropies are present, the effective dynamics governed by the above Hamilton's equations leads to a bounded energy density with its maximum coinciding with the maximum energy density in the isotropic $k=0$ case. Due to mathematical complexity a modified generalized Friedmann equation in terms of the energy density and the anisotropic shear is currently not available for the Bianchi-I model, but analyses based on the Hamilton's equations reveal the upper bounds on mean Hubble rate and the anisotropic shear \cite{Gupt:2011jh} which turn out to be  $H_\mathrm{max}=1/(2\gamma\lambda)$ and $\sigma^2_\mathrm{max}=10.125/(3\gamma^2\lambda^2)$.
Interestingly, numerical simulations for different types of matter and potentials in the Bianchi-I model reveal a novel seemingly universal parabolic relationship between the energy density and shear at the bounce which can shed important insights on the generalized Friedmann equation with quantum geometric modifications \cite{McNamara:2022dmf}. Further, these investigations reveal that the anisotropic shear at the bounce reaches its maximum value when the energy density at the bounce reaches approximately half of its maximum allowed value.

The evolution of the loop quantized Bianchi-I universe turns out to be non-singular for various types of matter with a bounce taking place at the minimal mean scale factor, see for example Fig. \ref{scalefactors} for a Bianchi-I universe filled with a massive scalar field. In particular, it can be shown that all the curvature invariants are bounded and the strong singularities are resolved throughout the non-singular evolution of the Bianchi-I universe as long as the null energy condition is satisfied \cite{Singh:2011gp}. The geometric structures of the bounces are also richer than those in the isotropic case where only the point-like bounce appears. In contrast, in the presence of anisotropies, the bounces can have barrel-, pancake- and cigar-like structures which can be obtained from a backward/forward evolution of the effective dynamics in the expanding/contracting phase using the same initial conditions in the classical regime that can lead to the singularities of the same type in GR. Therefore, these bounces can be regarded as the finite versions of the different types of the classical singularities which are resolved due to the underlying discreteness of the quantum geometry. Moreover, the quantum geometric effects also play the role of bridging different geometric structures in the pre- and post-bounce phases, leading to a Kasner transition that is impossible in the classical theory \cite{Gupt:2012vi, Singh:2016jsa,Wilson-Ewing:2017vju}. These Kasner transitions respect specific selection rules depending on the matter content of the universe. For example, in a Bianchi-I universe filled with the dust and radiation, the structure of the universe when approaching the bounce must be cigar-like on at least one side of the bounce. Therefore, in such a universe there is no pancake-pancake transition across the bounce. On the other hand, in a Bianchi-I universe filled with the stiff matter, there is no pancake like structure on either side of the bounce and depending on the anisotropic parameters, certain types of the transitions are more favored over the other types \cite{Gupt:2012vi}. Finally, the inflationary and ekpyrotic scenarios have been studied in the setting of loop quantized Bianchi-I models. In particular,
inflationary attractors for $\phi^2$ inflation have been found  in the Bianchi-I model in LQC and it has been shown that there exists a quantum bounce in the past evolution of the inflationary
trajectories and anisotropies remain bounded throughout the evolution \cite{Gupt:2013swa}. Similarly, non-singular ekpyrotic model has been constructed using effective dynamics of Bianchi-I LQC \cite{Cailleteau:2009fv}.

\subsection{Fock quantized inhomogeneities in polymer background: polarized Gowdy models}

In addition to the cosmological spacetimes in the mini-superspace characterized by  finite degrees of freedom, loop quantization has also been applied to those in the midi-superspace where inhomogeneities are also present as infinite degrees of freedom of the system \cite{Martin-Benito:2008eza,MenaMarugan:2009dp,Brizuela:2009nk,Garay:2010sk,Martin-Benito:2010vep,Martin-Benito:2010dge}. The motivation to study these spacetimes is to check the robustness of the physical implications, such as the singularity resolution, resulting from the quantum geometry effects as well as to investigate the impacts of the continuous degrees of freedom on the qualitative dynamics of LQC universe. One of the examples of this type is the Gowdy models which describe globally hyperbolic spacetimes with two spacelike commuting Killing vector fields and compact spatial sections \cite{Gowdy:1973mu}. The simplest one of the Gowdy models is  the linearly polarized Gowdy $T^3$ model. Since the Killing vector fields are also orthogonal to the hypersurface in this case, the spatial section $T^3$ can be described by coordinates $(\zeta,\eta,\delta)$ with $\zeta,\eta,\delta\in S^1$ and thus  two Killing vector fields are $\partial_\eta$ and $\partial_\delta$. Since the metric components which only depend on $t$ and $\zeta$ are periodic in $\zeta$, they can be expanded in the Fourier series in the $\zeta$-momentum space.  Using the symmetry of the system, one can completely fix the gauge freedom associated with the diffeomorphism constraints in $\eta$ and $\delta$ \cite{MenaMarugan:2009dp}, leaving alone two global constraints: the diffeomorphism constraint $C_\zeta$ which generates translation in  $\zeta$ and the Hamiltonian constraint $C_H$ which consists of a homogeneous part $C_\mathrm{hom}$ and an inhomogeneous part $C_\mathrm{inh}$.

The loop quantization of the Gowdy $T^3$ model adopts a hybrid quantization method, where the homogeneous sector is loop quantized while the inhomogeneous degrees of freedom are Fock quantized. A complete loop quantization requires a polymer quantization of the inhomogeneities which remains to be achieved. In the hybrid approach, the classical diffeomorphism and the Hamiltonian constraints are expressed in terms of the variables adapted to this quantization scheme. In particular, since the homogeneous sector of the Gowdy $T^3$ model  is equivalent to the vacuum Bianchi-I spacetime, it is appropriate to be parameterized by the  Ashtekar-Barbero connection $c_i$, with $i=\zeta,\eta,\delta$, and the corresponding densitized triads $p_i$ introduced in Sec. \ref{Hamiltonian_Bianchi-I}. As for the inhomogeneous degrees of freedom, Fock quantization scheme prefers the use of the creation and annihilation variables $(a_m,a^*_m)$, where $m$ can be  any non-zero integer, standing  for the non-zero modes that can be naturally associated with a free massless scalar field. In terms of the chosen variables, the expressions of the  diffeomorphism and the Hamiltonian constraints can be obtained which are given explicitly in \cite{Garay:2010sk}. Then using the techniques described in the above subsection for Bianchi-I spacetime and promoting the creation and annihilation variables to their operator analogs, the Hamiltonian constraint operator, a complete set of the Dirac observables as well as the physical inner product can be constructed precisely following the same strategy as used in the loop quantization of the spatially-flat FLRW spacetime.  The annihilation of the physical states by the Hamiltonian constraint operator  also leads to a quantum difference equation with uniform steps in the volume. As a result, the superselection sector in the kinematic Hilbert space of the Bianchi-I spacetime is also retained in the Gowdy $T^3$ model. Moreover, the zero volume states in the homogeneous sector turn out to be decoupled by the action of the Hamiltonian constraint. As a result, the singularity is resolved in the quantum theory.  Another interesting property of the quantized  Gowdy $T^3$ model is that its physical Hilbert space is a tensor product of the physical Hilbert space of the Bianchi-I model and a Fock space in the standard Fock quantization, implying that the standard quantum field theory is actually recovered in the framework of loop quantization. This provides a prototype for considering the perturbations over a loop quantized cosmological background. Finally, the Planck physics in the  Gowdy $T^3$ model has also been explored by using the effective dynamics of LQC which again confirms that the singularity is really replaced by the quantum bounce  and one of the effects of the inhomogeneities is to increase the bounce volume as compared with the LQC Bianchi-I spacetime \cite{Brizuela:2009nk}.

\section{Beyond standard LQC: incorporating additional elements from LQG}
\lb{sec:alternative_cosmological_model}
\renewcommand{\theequation}{7.\arabic{equation}}\setcounter{equation}{0}

So far we have discussed various results in different cosmological models, including in presence of anisotropies and inhomogeneities which show the robustness of singularity resolution in LQC. Let us here note that since LQC is not derived from LQG it is important to understand the way physics of Planck scale changes when we include more elements from LQG. The goal of this section is to understand two such inputs and the way they affect predictions in LQC. We overview two variant cosmological models for the spatially-flat FLRW universe that originate from a separate treatment of the Lorentzian term in the classical Hamiltonian constraint of GR. These models can be regarded as extensions of standard LQC due to different quantization prescriptions. As discussed in Sec. \ref{sec:LQC_in_spatially_flat_universe}, in the spatially-flat LQC, due to underlying symmetry the Euclidean and Lorentzian terms are combined before quantization. But how do the quantum constraint and its predictions change when we treat them separately? An independent treatment of the Lorentzian term using the Thiemann regularization was first discussed in \cite{Yang:2009fp} where the corresponding quantum Hamiltonian constraints resulting from two different ways of regularization of the Lorentzian term were derived. In the following, these two variants are called mLQC-I and mLQC-II, meaning the modified LQC model I and II, following the convention in \cite{Li:2018opr}. Later, mLQC-I was rediscovered in \cite{Dapor:2017rwv} from computing the expectation value of the quantum Hamiltonian constraint in LQG by  using the complexifier coherent states developed by Thiemann and Winkler \cite{Thiemann:2000bw,Thiemann:2000bx}.
 Different from standard LQC, the quantum difference equations in mLQC-I/II turn out to be 4th order difference equations \cite{Saini:2018tto} which lead to more complicated structure of the effective Hamiltonian constraints. This also results in a generic resolution of strong curvature singularities \cite{Saini:2019tem}.  The correct forms of the modified Friedmann and Raychaudhuri  equations in mLQC-I were found in \cite{Li:2018opr} implying an emergent quasi de-Sitter phase in the contracting branch \cite{Assanioussi:2018hee,Assanioussi:2019iye} with a rescaled Newton's constant \cite{Li:2018opr}.  On the other hand, the dynamics of mLQC-II turns out to be very similar to standard LQC with a symmetric bounce and modified maximum energy density. The phenomenological implications of these two models are extensively discussed in the literature, including both the background dynamics and the linear perturbations \cite{Li:2018fco,Li:2019ipm,Agullo:2018wbf,Garcia-Quismondo:2019dwa,Li:2019qzr,Li:2020mfi,Li:2021mop}. In the following, we start with the effective dynamics of mLQC-I/II and then briefly mention their implications on the inflationary scenario.

\subsection{The effective dynamics of mLQC-I/II}

Similar to the derivation of the effective dynamics in LQC, using the sharply peaked Gaussian coherent states, one can obtain the effective Hamiltonian constraint of mLQC-I/II by computing the expectation value of the Hamiltonian constraint operator in each model. In mLQC-I, the effective Hamiltonian constraint takes the form \cite{Yang:2009fp,Dapor:2017rwv}
\bqn
\lb{ham-mLQC-I}
\mathcal {H}^{\scriptscriptstyle{\mathrm{I}}}=\frac{3v}{8\pi G\lambda^2}\Big\{\sin^2(\lambda b)-\frac{(\gamma^2+1)\sin^2\left(2\lambda b\right)}{4\gamma^2}\Big\}+\mathcal{H}_M ~.
\eqn
With the help of the Poisson bracket $\{b,v\}=4\pi G\gamma$, it is straightforward to derive the Hamilton's equations of the system. The novel property of mLQC-I shows up when we try to derive the modified Friedmann equation from the Hamilton's equation of the volume. From the Hamiltonian constraint, one can find two branches, i.e. the $b_\pm$ branches with the relations
\bq
\lb{Hcd}
\sin^2(\lambda b_{\pm})= \frac{1\pm\sqrt{1-\rho/\rho^{\scriptscriptstyle{\mathrm{I}}}_c}}{2(\gamma^2+1)},
\eq
where $\rho^{\scriptscriptstyle{\mathrm{I}}}_c \equiv3/[32\pi G \lambda^2\gamma^2(\gamma^2+1)]$ is the maximum energy density that can be attained in the model. The difference between these two branches can be seen directly from the above relation. In particular,  when the energy density tends to vanish, $b_-$ goes to zero while $b_+$ approaches a non-zero value. This implies that the classical limit can only be recovered in the $b_-$ branch. A detailed analysis on the modified Friedmann equation confirms this intuition.  Substituting the relation (\ref{Hcd}) into the  Hamilton's equation of the volume and then squaring it, one can obtain the modified Friedmann equation for each branch \cite{Li:2018opr}. For the $b_-$ branch, the modified Friedmann equation turns out to be
\bq
H^2_-=\frac{8\pi G \rho}{3}\left(1-\frac{\rho}{\rho^{\scriptscriptstyle{\mathrm{I}}}_c}\right)\Bigg[1  +\frac{\gamma^2}{\gamma^2+1}\left(\frac{\sqrt{\rho/\rho^{\scriptscriptstyle{\mathrm{I}}}_c}}{1 +\sqrt{1-\rho/\rho^{\scriptscriptstyle{\mathrm{I}}}_c}}\right)^2\Bigg],
\eq
which asymptotes to the classical Friedmann equation as $\rho\ll\rho^{\scriptscriptstyle{\mathrm{I}}}_c$.  The $b_-$ branch describes the expanding phase of the universe in mLQC-I. In the backward evolution of the universe, the volume decreases continuously with an increasing energy density. When $\rho$ increases to $\rho^{\scriptscriptstyle{\mathrm{I}}}_c$, a quantum bounce takes place irrespective of the matter content. In this process, similar to the standard LQC, when the energy density increases to a threshold value in the Planck regime (less than $\rho^{\scriptscriptstyle{\mathrm{I}}}_c$), the universe enters into a super-inflationary phase in which $\dot H>0$. In mLQC-I, this threshold value of the energy density depends on the Barbero-Immirzi parameter $\gamma$ and its numerical value is about $\rho_s^-\approx 0.503 \rho^{\scriptscriptstyle{\mathrm{I}}}_c$, with $\gamma\approx0.2375$. On the other hand, the contracting phase of the universe in mLQC-I is described by the $b_+$ branch in which the modified Friedmann equation is given by
\bq
H^2_+ =\frac{8\pi G_\alpha  \rho_\Lambda}{3}\left(1-\frac{\rho}{\rho^{\scriptscriptstyle{\mathrm{I}}}_c}\right)\left[1+\left(\frac{1-2\gamma^2+\sqrt{1-\rho/\rho^{\scriptscriptstyle{\mathrm{I}}}_c}}{4\gamma^2\left(1+\sqrt{1-\rho/\rho^{\scriptscriptstyle{\mathrm{I}}}_c}\right)}\right)\frac{\rho}{\rho^{\scriptscriptstyle{\mathrm{I}}}_c}\right],
\eq
where the rescaled Newton's constant $G_\alpha\equiv G ({1-5\gamma^2})/({\gamma^2+1})\approx 0.680 G$ and the emergent cosmological constant $\rho_\Lambda \equiv{3}/[{8\pi G\alpha \lambda^2(1+\gamma^2)^2}]\approx0.0304$. Hence, this branch is characterized by an emergent quasi de Sitter phase with a rescaled Newton's constant at zero energy density. When $\rho\rightarrow0$, the Hubble rate $H^2\rightarrow \frac{8\pi G\alpha  \rho_\Lambda}{3}\approx 0.173$ which is almost of Planck scale.  Thus, the evolution of the universe in mLQC-I is asymmetric with respect to the quantum bounce.\footnote{If instead of working with cosmic time, one chooses massless scalar field as a clock, one reaches infinite volume at a finite value of this clock. With such a clock, quantum extensions in the physical Hilbert space have been explored assuming a massless scalar field as the only matter content in Ref. \cite{Assanioussi:2019iye}. } Surprisingly, in the $b_+$ branch, the Hubble rate becomes vanishing at the minimum energy density $\rho_\mathrm{min}=-3/(8\pi G\lambda^2)\approx-0.023$ which is a negative number.  As a result, a necessary condition for realization of a cyclic universe in mLQC-I is the violation of the weak energy condition \cite{Li:2021fmu}. The super-inflationary phase in the $b_+$ branch  begins when the energy density reaches a different threshold value   $\rho_s^+ \approx 0.377 \rho^{\scriptscriptstyle{\mathrm{I}}}_c$.
Therefore, the super-inflationary regime in mLQC-I is also asymmetric with respect to the quantum bounce. Considering all the remarkable  properties of its contracting branch, mLQC-I serves as one of the concrete examples  where different quantization prescriptions  can result in qualitatively distinct  quantum dynamics.
In comparison to the novel properties of mLQC-I, the qualitative dynamics of mLQC-II resembles that of the standard LQC. In particular, the evolution of the universe filled with a massless field is also symmetric with respect to the bounce.  The effective dynamics of mLQC-II is governed by the modified Friedmann equation  \cite{Li:2018fco}
\bq
H^2=\frac{16\pi G \rho}{3}\left(1-\frac{\rho}{\rho^{\scriptscriptstyle{\mathrm{II}}}_c}\right) \left(\frac{1+4\gamma^2(\gamma^2+1)\rho/\rho^{\scriptscriptstyle{\mathrm{II}}}_c}{1+2\gamma^2\rho/\rho^{\scriptscriptstyle{\mathrm{II}}}_c+\sqrt{1+4\gamma^2(1+\gamma^2)\rho/\rho^{\scriptscriptstyle{\mathrm{II}}}_c}}\right),
\eq
with  the maximum energy density $\rho^{\scriptscriptstyle{\mathrm{II}}}_c=3\left(\gamma^2+1\right)/2\pi G \gamma^2\lambda^2$, for both contracting and expanding branches. The quantum bounce takes place generically  at the maximum energy density $\rho=\rho_c^{{\scriptscriptstyle{\mathrm{II}}}}$ and the  super-inflation occurs for $\rho \geq 0.513\rho_c^{{\scriptscriptstyle{\mathrm{II}}}}$. Although  there is no qualitative distinctions between mLQC-II and standard LQC as discovered so far,  it still implies that quantization ambiguities can result in a different maximum energy density and durations of the super-inflationary phase in the Planck regime, that is, except for some generic features such as the resolution of the big bang singularity by a quantum bounce,  the details of the quantum dynamics are also closely related with the specific quantization prescriptions.

As the second part of this section, let us briefly summarize the results of the inflationary background dynamics  in mLQC-I/II. Similar to standard LQC, the embedding of the inflationary paradigm into mLQC-I/II has been extensively studied in the literature \cite{Li:2018fco,Li:2019ipm}. Taking into account an inflationary potential, it has been found that the inflationary phase is also a local attractor of the qualitative dynamics in each model. Detailed analysis has been done for the $\phi^2$ potential,  the fractional monodromy potential, the Starobinsky potential, the non-minimal Higgs potential and the exponential potential. In particular,  the numerical and analytical results of the background evolution with the chaotic and the Starobinsky potentials are available in mLQC-I/II. It has been shown explicitly that similar to  standard LQC, for the kinetic-energy-dominated bounce, the evolution of the universe by the end of the inflationary (slow-roll) phase in the expanding  branch  can be divided into three distinctive stages: the bouncing phase, the transition phase and the slow-roll phase. Following the approach used in LQC,  the probability of the occurrence of the desired slow-roll phase is also estimated  in mLQC-I/II:   one can find that the probability for the desired slow-roll not to happen for a $\phi^2$ potential is
$P^{\scriptscriptstyle{\mathrm{I}}}(\mathrm{not ~realized}) \lesssim1.12\times 10^{-5}$ in mLQC-I and
$P^{\scriptscriptstyle{\mathrm{II}}}(\text{not realized})\lesssim 2.62\times 10^{-6}$ in mLQC-II.

Though inflationary paradigm is compatible with both mLQC-I and mLQC-II, the same can not be said about the alternatives of inflation based on cyclic models. A surprising result is that ekpyrotic models are incompatible with mLQC-I \cite{Li:2021fmu}. Note that this is not because mLQC-I does not result in singularity resolution rather the problem comes from the lack of recollapse in the classical regime after one cycle as a result multiple cycles can not occur. Given the conventional wisdom that cyclic models are easy to construct with a non-singular bounce if one adds matter which causes a recollapse at late times, this result is indeed counter-intuitive but can be easily understood. As we have discussed, while standard LQC results in a classical regime in both pre-bounce and post-bounce epochs at large volumes, the same can not be said for mLQC-I where the  pre-bounce regime has a large effective cosmological constant phase originating from quantum geometry. It turns out that every bounce in mLQC-I switches the branch from $b_+$ to $b_-$ (or vice-versa). In the case of a model where there is a recollapse it turns out that the asymptotic past and future branches, after the classical turnaround and the second bounce, are dictated by the $b_+$ branch. Therefore, unlike a model with single bounce in mLQC-I  which connects $b_+$ and $b_-$ branch, the additional classical turnaround and the subsequent bounce converts the future asymptotic branch after the second bounce  to a $b_+$ branch.  As discussed earlier the $b_+$ branch does not correspond to the classical spacetime with low curvature, which forbids another recollapse and a subsequent bounce. Hence, even in presence of a non-singular bounce a cyclic evolution with more than one cycle is not possible in mLQC-I unless one chooses a very large negative potential  which overcomes Planckian effective cosmological constant. On the other hand, given the symmetric nature of bounce cyclic models are as much compatible with mLQC-II as in standard LQC. An important question is whether this result implies that cyclic models are inconsistent with LQG. While this remains an open question, it is indeed true that highly asymmetric bounces leading to only one side of evolution with low spacetime curvatures are  not just confined to mLQC-I but also occur in certain loop quantizations of Kantowski-Sachs spacetime \cite{Dadhich:2015ora}. One may conjecture that if LQG results in a evolution characterized by such an asymmetric bounce it would result in incompatibility with cyclic models.

\section{Summary and Outlook}
\lb{sec:conclusions}
\renewcommand{\theequation}{8.\arabic{equation}}\setcounter{equation}{0}

~~~~~The development of LQC in the last two decades has made itself a prototype to exemplify how the underlying discrete quantum geometry effects can resolve strong curvature singularities intrinsic in the classical theory in various cosmological settings and meanwhile provide an insightful picture of the physics beyond the standard cosmology, especially of the quantum evolution of the universe in the Planck regime. Featured by the background independent formulation of the quantum theory, LQC makes use of the Ashtekar-Barbero variables tailored to symmetry reduced spacetimes to construct the quantum Hamiltonian constraint operator as well as relevant physical observables. Its kinematic Hilbert  space is made up of essentially discrete quantum states which provide a quantum representation that is unitarily inequivalent to the Schr\"odinger representation. Another key element in the formulation is to introduce a minimal non-zero area gap to build up the non-local curvature operator. This becomes a vital step to yield a physically viable loop quantization which on one hand results in a non-singular evolution of the universe in the Planck regime and on the other hand asymptotes to the right classical limit in the low curvature regime. Its unique quantum representation makes itself distinct as compared with the Wheeler-Dewitt theory in the sense that LQC can only be well-approximated by the Wheeler-Dewitt theory far away from the Planck regime.

The resolution of big bang singularity in LQC was first discovered in the simplest case of a homogeneous and isotropic FLRW universe filled with a massless scalar field where the numerical simulations of the quantum difference equation revealed a quantum bounce in the Planck regime \cite{Ashtekar:2006wn}. Later, the robustness of the singularity resolution as well as the existence of the quantum bounce has been tested in more complicated contexts, such as spacetimes with a cosmological constant, non-vanishing spatial curvature, anisotropies, continuous degrees of freedom \textcolor{black}{and even in some modified gravity theories \cite{Zhang:2012em,Song:2020pqm}}. Supported by the numerical simulations of the quantum difference equation, the effective description of the quantum evolution of the LQC universe plays an important role in achieving an analytical understanding of the singularity resolution in the Planck regime in various types of loop quantized spacetimes \cite{Singh:2009mz,Singh:2014fsy}. Moreover, using the effective dynamics of LQC, well-known scenarios in the standard cosmology, such as the inflationary paradigm, the ekpyrotic and matter bounce scenarios, have been extensively studied with a purpose of providing a ultra-violet complete description of these scenarios in the context of the LQC universe. In addition to these topics which we discussed earlier, quantum geometric effects in LQC have also been applied in order to obtain non-singular evolution in string motivated scenarios such as pre-big bang cosmology \cite{DeRisi:2007lpg} and multiverses \cite{Gupt:2013poa,Garriga:2013cix}.
The impact of the quantum geometry effects on the dynamical evolution of the universe in all these scenarios as well as their detectable characteristic signals become the central subjects to investigate at the phenomenological level. Among all of these progresses, it is worthwhile to point out that the inflationary phase is found to be a local attractor in the LQC universe with/without anisotropies when the initial conditions of the dynamical trajectories are set in the Planck regime and the probability for the occurrence of inflation can also be computed explicitly due to the availability of a distinguished physical measure of the parameter space right at the quantum bounce in a homogeneous and isotropic LQC universe.

In addition to coupling to the massless scalar field, the quantum theory for a homogeneous and isotropic LQC universe filled with a massive scalar field can also be constructed with a  similar procedure \cite{Giesel:2020raf}. The  novel feature in presence of the potential is that the scalar field can no longer serve as a global clock. As a result, some other reference fields are required to play the role of the clocks. One of the strategies is that the quantization can be implemented in the reduced phase space formulated in the relational formalism in which the Dirac observables are explicitly constructed before quantization is carried out. To deparametrize the theory, one needs to introduce some matter fields, such as the dust fields or the Klein-Gordon scalar fields, which serve as the reference fields with respect to which physical observables are constructed explicitly.  For the Gaussian and the Brown-Kucha\v r dust fields, the resulting physical Hamiltonian in the reduced phase space takes the same form as its analog in the classical phase space with classical phase space variables replaced by their observable counterparts in the reduced phase space. Moreover, the physical Hamiltonian is no longer constrained to vanish. One can then construct the quantum theory using the techniques well developed in LQC and  it turns out that the impacts of the different reference fields on the quantum dynamics of the LQC universe can be tuned as small as possible with a careful choice of the initial energy density of the reference fields. This line of research is aimed at solving the problem of time in a general setting of LQC beyond the limit of the massless scalar field and it helps justify the effective dynamics of LQC with a massive scalar field which is used in prevail in the literature for a phenomenological investigation of  the cosmological implications of the LQC universe.  Apart from the treatment of the massive scalar field, we also believe that a consistent loop quantization of both the geometric and matter degrees of freedom on the same footing is also important in achieving a general picture of the quantum dynamics of the LQC universe in the Planck regime.

Meanwhile recent progress has been made in the direction of exploring the alternative quantization prescriptions other than the one used in standard LQC. The motivation mainly originates from endeavors to unravel the relationship between LQC and LQG. Due to the non-commutativity between quantization and symmetry reduction, it is a long-standing question to ask that how many features LQC has really inherited from the cosmological sector of full LQG. The bottom-to-top approach to addressing this concern has led to alternative LQC models, such as mLQC-I/II. Although some generic features, such as singularity resolution and the existence of a quantum bounce which connects a contracting phase with an expanding one, still remain true in these modified LQC models, they also exhibit novel features of the quantum dynamics even in the simplest setting of a homogeneous and isotropic FLRW universe and provide some evidence to show that the quantum dynamics in the Planck regime can have much richer structures than originally proposed in standard LQC. In particular, the universe can evolve asymmetrically  with respect to the quantum bounce and \textcolor{black}{the evolution of the contracting phase can also proceed in a de Sitter phase with a Planck-scale cosmological constant. The emergent de Sitter phase in the contracting phase has also been observed in other alternative quantization of the Hamiltonian constraint constructed from the Chern-Simons action \cite{Yang:2019ujs,Yang:2020eby} where the big bang singularity is once again resolved in the case of the spatially-flat FLRW universe. Thus, }whether the classical universe can be recovered on the other side of the bounce is now in  question.  \textcolor{black}{Although a bouncing evolution can be shown as obtained from a specific model of loop quantum gravity \cite{Zhang:2019dgi}}, the above results imply that standard LQC may not capture all of the features of the quantum cosmological sector of LQG. In addition to the modified LQC models, other strategies to investigate the connection between LQC and LQG include the quantum-reduced loop quantum gravity  \cite{Alesci:2013xd}, the group theory cosmology \cite{Gielen:2013kla,Gielen:2013naa} and the path integral approach \cite{Han:2019feb}. It is remarkable to note that although the improved dynamics with the $\bar \mu$ scheme was initially found in the homogeneous and isotropic FLRW LQC universe, its validity has extended to the other spacetimes and quantization prescriptions. Moreover, the $\bar \mu$ scheme is also found in other approaches such as   the quantum-reduced loop quantum gravity  and the path integral approach. Therefore,  %and the uniqueness of the
the $\bar \mu$ scheme, which is the only viable prescription known so far in homogeneous and isotropic cosmological models of LQC, seems to be consistent even in a more general setting.
%some fundamental physical principles which is yet to be discovered.

Finally, although in this chapter we have mainly focused on the cosmological spacetimes, it is important to note that loop quantization is not only restricted to the FLRW universe. Numerous works have been done to apply loop quantization to the spherically symmetric spacetimes for the purpose of exploring the role of the non-perturbative quantum geometry effects in resolving the strong singularity encountered in the classical black hole spacetimes. Depending on the degrees of freedom that are to be quantized, the models of LQG black holes can be classified into  two main categories. The models in the first category are quantized in the mini-superspace where symmetry reduction and homogeneity are assumed first at the classical level. In this case, the isometry between the Schwarzschild interior and the Kantowski-Sachs spacetime in cosmology  is usually used and one only needs to deal with finite degrees of freedom \cite{Ashtekar:2005qt,Boehmer:2007ket,Corichi:2015xia, Olmedo:2017lvt, Ashtekar:2018lag,Bodendorfer:2019cyv}. Here quantum difference equation has much richer structure than the isotropic LQC whose simulations also indicate singularity resolution \cite{Yonika:2017qgo}. On the other hand, the models in the second category only appeals to the spherical symmetry and one needs to deals with the infinite degrees of freedom \cite{Bojowald:2005cb,Chiou:2012pg,Gambini:2013hna,Gambini:2020nsf,Kelly:2020lec,Han:2020uhb} which make them similar to the  Gowdy models in cosmology. In addition to the vacuum spacetimes, investigations have also been extended to the cases which include some matter fields, such as the scalar fields and the dust fields \cite{Han:2020uhb,Bojowald:2005qw,Gambini:2009ie,Bambi:2013caa,Benitez:2020szx,Giesel:2022rxi}. In all these cases, similar to the big bang singularity in cosmology, the curvature singularity at the center of black hole is resolved and replaced by a bounce. However, the physics beyond the bounce is model-dependent. The black hole can be glued to a white hole with the same/different mass or Nariai universe resulting from an emergent cosmological constant. Research in the LQG black holes is still full of actively ongoing subjects and more astounding discoveries are expected in the near future.

\section*{Acknowledgement}
 We are grateful to many colleagues over the years for various stimulating discussions on singularity resolution in LQC in particular Abhay Ashtekar,  Alejandro Corichi, David Craig, Naresh Dadhich, Peter Diener, Kristina Giesel, Brajesh Gupt, Anton Joe, Andrzej Kr\'olak, Klaus Liegener, Meenakshi McNamara, Javier Olmedo, Tomasz Pawlowski,   Sahil Saini,  Kevin Vandersloot, Alexander Vilenkin, Anzhong Wang and Edward Wilson-Ewing. We thank Meenakshi McNamara for help with Fig 2. B.-F. Li acknowledges support by the National Natural Science Foundation of China (NNSFC) with the grant No. 12005186. P.S. is supported by NSF grants PHY-1912274 and PHY-2110207.

\end{document}